\documentclass[pre,reprint,twocolumn,superscriptaddress,floatfix,aps]{revtex4-2}
\usepackage{amsmath, amssymb, bm}
\usepackage{graphicx}
\usepackage{physics}
\usepackage[T1]{fontenc}
\usepackage{comment}
\usepackage{float}
\usepackage{mathtools}
\usepackage[normalem]{ulem}
\usepackage{hyperref}
\hypersetup{colorlinks, 
 linkcolor=blue, %
 citecolor=blue, %
 urlcolor=blue}
\usepackage{color}

\begin{document}
\title{Synchronization in adaptive higher-order networks}

\author{Md Sayeed Anwar}
\email{sayeedanwar447@gmail.com}
\affiliation{Physics and Applied Mathematics Unit, Indian Statistical Institute, 203 B. T. Road, Kolkata 700108, India}
\affiliation{Department of Mathematics and Namur Institute for Complex Systems, naXys, University of Namur, 2 rue Grafé, Namur B5000,
Belgium}

\author{S. Nirmala Jenifer}
\affiliation{Department of Physics, Bharathidasan University, Tiruchirappalli 620024, Tamil Nadu, India}

\author{Paulsamy Muruganandam}
\affiliation{Department of Physics, Bharathidasan University, Tiruchirappalli 620024, Tamil Nadu, India}

\author{Dibakar Ghosh}
\affiliation{Physics and Applied Mathematics Unit, Indian Statistical Institute, 203 B. T. Road, Kolkata 700108, India}

\author{Timoteo Carletti}
\email{timoteo.carletti@unamur.be}
\affiliation{Department of Mathematics and Namur Institute for Complex Systems, naXys, University of Namur, 2 rue Grafé, Namur B5000,
Belgium}

\begin{abstract}
Many natural and human-made complex systems feature group interactions that adapt over time in response to their dynamic states. However, most of the existing adaptive network models fall short of capturing these group dynamics, as they focus solely on pairwise interactions. In this study, we employ adaptive higher-order networks to describe these systems by proposing a general framework incorporating both adaptivity and group interactions. We demonstrate that global synchronization can exist in those complex structures, and we provide the necessary conditions for the emergence of a stable synchronous state. Additionally, we analyzed some relevant settings, and we showed that the necessary condition is strongly related to the master stability equation, allowing to separate the dynamical and structural properties. We illustrate our theoretical findings through examples involving adaptive higher-order networks of coupled generalized Kuramoto oscillators with phase lag. We also show that the interplay of group interactions and adaptive connectivity results in the formation of stability regions that can induce transitions between synchronization and desynchronization. 
\end{abstract}

\maketitle

\section{Introduction}
\label{sec:intro}

In the realm of complex systems, synchronization is a captivating and widespread phenomenon where coupled systems spontaneously self-organize by displaying a coordinated behavior \cite{synchronization1, arenasreview}. This phenomenon, emerging in both natural and artificial systems, has long intrigued scientists seeking to understand its underlying principles \cite{boccaletti2018synchronization}.

Network science has emerged as a powerful framework for studying synchronization, where interconnected nonlinear oscillators are represented as nodes, and their interactions as pairwise links \cite{barabasibook}. However, traditional network models have limitations when applied to many natural and human-made systems, such as brain networks \cite{ bassett2011dynamic} and social networks \cite{wasserman1994social}, where connections between individual units are dynamic and evolve over time. To address this issue, the concept of networks has been generalized as to include temporally evolving connectivity topology \cite{holme2012temporal}. A particularly intriguing class of these generalized network structures is given by the adaptive networks, where the temporal evolution of the network structure is intricately linked to the dynamical state of its nodes, leading to the coevolution of both the network topology and its individual components \cite{sayed2014adaptive, gross2009adaptive, berner2023adaptive,sawicki2023perspectives}. For example, synaptic connections between neurons adjust based on the relative timing of neuronal spikes \cite{markram1997regulation, abbott2000synaptic, meisel2009adaptive, lucken2016noise}. Likewise, in certain chemical systems, reaction rates dynamically adapt according to the system variables \cite{jain2001model}. Activity-dependent plasticity is also prevalent in epidemics and various biological or social systems \cite{gross2006epidemic, gross2008adaptive}. Researchers have investigated synchronization within adaptive networks \cite{berner2021desynchronization, berner2021synchronization, thiele2023asymmetric, berner2020birth, manoranjani2023phase, lodi2024patterns, gutierrez2011emerging, zhang2015explosive}, as well as in conventional time-varying networks \cite{ghosh2022synchronized, kohar2014synchronization, stilwell2006sufficient, CarlettiFanelli2022, intra2}, where the temporal evolution of connections is predefined and independent of node dynamics.

Another limitation of traditional network models is their focus on pairwise interactions, which fails to capture the complexity of many real-world systems. To overcome this limitation, researchers have highlighted the importance of higher-order structures that go beyond pairwise links by allowing for simultaneous interactions among multiple agents \cite{battiston2020networks, battiston2021physics, boccaletti2023structure, majhi2022dynamics}. Higher-order structures, such as simplicial complexes \cite{bianconi2021higher} and hypergraphs \cite{berge1973graphs}, offer a more nuanced understanding of complex systems and have revealed new features in various dynamical processes, including epidemics \cite{iacopini2019simplicial}, random walks \cite{CarlettiEtAl2020, schaub2020random}, consensus \cite{schaub2}, pattern formation \cite{muolo2023turing, gao2023turing}, synchronization \cite{carlettifanellinicoletti2020, skardal2020higher, simplicialsync4, carletti2023global, anwar2022intralayer, anwar2022stability, jenifer2024synchronizability, gambuzza2021stability, anwar2022synchronization}, swarmalation \cite{anwar2024collective} and more \cite{ghosh2023first, anwar2024self}.

Despite these advances, the current frameworks still fall short in describing systems with both adaptivity and higher-order interactions. For instance, in neuronal networks, a group of neurons interact simultaneously \cite{tlaie2019high, amari2003synchronous} and also, the synaptic connectivity between them depends on the neuronal spike timing \cite{abbott2000synaptic, meisel2009adaptive, lucken2016noise}. In this work, we thus aim to close this gap by exploring the interplay between higher-order and adaptive interactions by proposing a general framework for studying synchronization involving nonlinear oscillators coupled via adaptive higher-order network structures. To the best of our knowledge, only a few recent studies have explored the adaptive nature of higher-order interactions and their effects on the synchronization process \cite{kachhvah2022first, rajwani2023tiered, dutta2024transition, emelianova2024adaptation}. For instance, it has been shown that group interactions in adaptive networks can influence synchronization transitions within the Kuramoto model \cite{rajwani2023tiered, dutta2024transition}. Let us observe that these studies have restricted their focus on the Kuramoto model and have been limited to examining only the routes to transitions between synchronization and desynchronization, such as continuous or abrupt transitions.

In this work, we overcome the limitations of the above-cited results and introduce a broad framework for studying dynamical systems within adaptive higher-order networks. Specifically, we consider a finite ensemble of generic (but identical) dynamical systems anchored to the nodes of an adaptive higher-order network, where the interactions are governed by general coupling functions. For simplicity, we focus on simultaneous interactions up to three bodies, but the proposed framework can be easily extended as to accommodate for interactions of any order. Within such context, we examine the synchronization, specifically global synchronization in adaptive higher-order networks. This approach extends the study by Berner et al. \cite{berner2021desynchronization} on synchronization in adaptive networks to the domain of higher-order networks. Given the existence of a synchronous solution, we derive the necessary conditions for its stability in two scenarios: first, when only the pairwise connections adapt according to node dynamics while higher-order connections remain fixed, and second, when both pairwise and higher-order connections adapt based on node dynamics. We considered a couple of relevant higher-order structures, all-to-all and ring-like, and we shown that the analytical stability conditions we derived, resemble the Master Stability Equation (MSE) approach \cite{fujisaka1983stability,msf}. This is a method originally developed for static pairwise networks, and since then extended to various complex networks, including time-varying networks \cite{CarlettiFanelli2022,stilwell2006sufficient, anwar2024global} and static higher-order networks \cite{gambuzza2021stability, anwar2022stability}. To validate the analytic derivations we obtained, we use the generalized Kuramoto model with phase lag defined on adaptive higher-order networks. We show how the combined effect of adaptivity and higher-order interactions change the stability of the synchronous state. We also demonstrate that the interplay between adaptation and group interactions can lead to the formation of stability regions, which induce transitions between synchronization and desynchronization.

\section{Adaptive pairwise and static higher-order interactions}
\label{sec:nohoiadapt}

Let us first consider the case where both pairwise and higher-order interactions are present, but only the former can evolve in time as a function of the system state variables; the three-body terms are thus supposed to be given by a stationary tensor. More precisely, we are considering the system, 
\begin{align}
{\dot {\mathbf x}}_i = & {f}(\mathbf{x}_i) - \sigma_1 \sum_{j=1}^N a_{ij}^{(1)} k_{ij}^{(1)}(t) {g}^{(1)}(\mathbf{x}_i, \mathbf{x}_j) \notag \\ & -\sigma_2{\sum_{j=1}^N}{\sum_{p=1}^N} a_{ijp}^{(2)} k_{ijp}^{(2)} {g}^{(2)}(\mathbf{x}_i, \mathbf{x}_j, \mathbf{x}_p)\, , 
\label{eq:eq1}
\end{align}
where the smooth function ${f}$ (assumed to be same for all nodes) describes the dynamics of $d$-dimensional isolated nodes, $a_{ij}^{(1)}$ (resp. $a_{ijp}^{(2)}$), is the adjacency matrix (resp. adjacency $2$-tensor) encoding for the ``topological'' pairwise (resp. three-body) interactions, namely for the possible interactions among the constituting basic units despite their actual realization due to the adaptation of the structures. The time-varying matrix $k_{ij}^{(1)}(t)$ describes the adaptation of the weight links due to the time evolution of the state variables. The $2$-tensor $k_{ijp}^{(2)}$ encodes for the weights of the three-body interaction; let us observe that for the time being, we assume it to be constant and thus we could have merged it with $a_{ijp}^{(2)}$, we, however, preferred to keep them separate from each other to use a similar notation in the following section, where also the higher-order interactions can adapt to the system state evolution. Finally, the function ${g}^{(1)}$ (resp. ${g}^{(2)}$) encodes the coupling involving two-body (resp. three-body) and $\sigma_j$, $j=1, 2$, denote the corresponding coupling strengths.

As already stated, weights of pairwise interactions adapt because of the evolution of the system state \cite{berner2019multiclusters, berner2021desynchronization, Kasatkin2017}, more precisely, we assume for all $i, j\in\{1, 2, \dots, N\}$,
\begin{align}
\label{eq:evolkij}
\dot k_{ij}^{(1)} = -\epsilon[k_{ij}^{(1)} + a_{ij}^{(1)}h(\mathbf{x}_i-\mathbf{x}_j)]\, , 
\end{align}
for some scalar coupling function ${h}$ and $0<\epsilon \ll 1$ is a parameter that separates the time scales of the slow dynamics of the coupling strengths from the fast dynamics of the oscillatory system. On the other hand, we assume the strengths of the higher-order interactions to be fixed over time, 
\begin{align}
k_{ijp}^{(2)}=1 \quad \forall i, j, p\in\{1, 2, \dots, N\}\, .
\notag\end{align}

Without any further assumption on the coupling functions, we require the following to hold true for the synchronization to occur
\begin{enumerate}
\item $\sum\limits_{j=1}^{N} a_{ij}^{(1)}=r^{(1)}$, i.e., each node must participate in the same number of pairwise interactions.
\item $\sum\limits_{j=1}^{N}\sum\limits_{p=1}^{N} a_{ijp}^{(2)}=2r^{(2)}$, i.e., each node must participate in the same number of triadic interactions.
\end{enumerate}
Stated differently, we are assuming that the pairwise structure, $a_{ij}^{(1)}$, and the higher-order one, $a_{ijp}^{(2)}$, are regular. This assumption of constant degree is essential to enable synchronization for generic coupling functions.  

Let us observe that we can overcome the above limitation of constant degree by considering the coupling functions $g^{(1)}$ and $g^{(2)}$ to be synchronization noninvasive, i.e., $g^{(1)}(\mathbf{s}, \mathbf{s})=0$ and $g^{(2)}(\mathbf{s}, \mathbf{s}, \mathbf{s})=0$. By imposing the noninvasive condition on the coupling functions, we can accommodate any network topology to achieve global synchronization. A detailed study of this case will be provided in Appendix \ref{sec:noninvasive}. 

\subsection{Emergence of synchronization}
Based on the above, we can look for a global synchronous solution $\mathbf{s}(t)$ of the system described by Eqs.~\eqref{eq:eq1}~-~\eqref{eq:evolkij}, namely a solution independent from the nodes indexes, it then follows that the latter should satisfy
\begin{align}
& \dot{\mathbf{s}} =  {f}(\mathbf{s}) + \sigma_1 r^{(1)}h(0) {g}^{(1)}(\mathbf{s}, \mathbf{s}) -2\sigma_2r^{(2)} {g}^{(2)}(\mathbf{s}, \mathbf{s}, \mathbf{s}) \\
& k_{ij}^{(s)} =  -a_{ij}^{(1)}h(0) \, .
\end{align}
To analyze the stability of this synchronous state, we perform linear stability analysis; namely, we consider perturbations of the form $\mathbf{\xi}_i = \mathbf{x}_i - \mathbf{s} $ and $\chi_{ij} = k_{ij}^{(1)} - k_{ij}^{(s)}$, whose time evolution is obtained by solving
\begin{widetext}
\begin{align}
\dot{\mathbf \xi}_i = & D{f}(\mathbf{s})\xi_i - \sigma_1 \sum_{j=1}^N a_{ij}^{(1)} {g}^{(1)}(\mathbf{s}, \mathbf{s}) \chi_{ij} 
+ \sigma_1r^{(1)}h(0)[D_1{g}^{(1)}(\mathbf{s}, \mathbf{s}) + D_2{g}^{(1)}(\mathbf{s}, \mathbf{s})] \xi_i + \notag \\ & 
- \sigma_1 h(0) \sum_{j=1}^N L_{ij}^{(1)} D_2{g}^{(1)}(\mathbf{s}, \mathbf{s})\xi_{j} - 2\sigma_2r^{(2)} [D_1{g}^{(2)}(\mathbf{s}, \mathbf{s}, \mathbf{s}) \notag + D_2{g}^{(2)}(\mathbf{s}, \mathbf{s}, \mathbf{s}) + D_3{g}^{(2)}(\mathbf{s}, \mathbf{s}, \mathbf{s})] \xi_i + \notag \label{eq:vareqxi} \\ & 
+ \sigma_2\sum_{j=1}^N L_{ij}^{(2)}[D_2{g}^{(2)}(\mathbf{s}, \mathbf{s}, \mathbf{s}) + D_3{g}^{(2)}(\mathbf{s}, \mathbf{s}, \mathbf{s})]\xi_j \\ 
{\dot{\chi}}_{ij} = & -\epsilon \left( \chi_{ij} + a_{ij}^{(1)} [Dh(0)(\mathbf{ {\xi}}_i - \mathbf{ {\xi}}_j)]\right) \label{eq:vareqchiij}\, , 
\end{align}
\end{widetext}
where $\mathbf{L}^{(1)}$ and $\mathbf{L}^{(2)}$ are the Laplace matrices for pairwise and three-body interactions, namely \begin{align}
 \begin{array}{l}
 L^{(1)}_{ij}=\begin{cases}
 -a^{(1)}_{ij}, & i \neq j \\
 \sum\limits_{j=1}^{N} a^{(1)}_{ij}=r^{(1)}, & i=j 
 \end{cases} 
 \end{array}
\end{align}
and, 
\begin{align}
 \begin{array}{l}
 L^{(2)}_{ij}=\begin{cases}
 -\sum\limits_{k=1}^{N} a^{(2)}_{ijk}, & i \neq j \\
 \sum\limits_{j=1}^{N}\sum\limits_{k=1}^{N} a^{(2)}_{ijk}=2r^{(2)}, & i=j 
 \end{cases} 
 \end{array}
\end{align}
$D_i{g}^{(j)}$ denotes the Jacobian of the function ${g}^{(j)}$, $j=1, 2$, with respect to the $i$-th variables, $i=1, 2, 3$.

To write the previous equations into a more compact form, we assume to cast the $N\times N$ matrix $\chi_{ij}$ into a $N^2$-dimensional vector by stacking the rows on rows successively, i.e., $\pmb{\chi}=(\chi_{11}, \dots, \chi_{1N}, \chi_{21}, \dots, \chi_{2N}, \dots, \chi_{N1}, \dots, \chi_{NN})^{\top}$. 
We can thus rewrite Eqs.~\eqref{eq:vareqxi} and~\eqref{eq:vareqchiij} as follows,
\begin{align}
\label{pairvariational}
\begin{bmatrix}
 \dot{\pmb{\xi}} \\
 \dot{\pmb{\chi}}
\end{bmatrix} 
= 
\begin{bmatrix}
 \mathbf{S} & -\sigma_1 \mathbf{B}^{(1)} \otimes {g}^{(1)}(\mathbf{s}, \mathbf{s}) \\
 -\epsilon \mathbf{C}^{(1)} \otimes Dh(0) & -\epsilon \mathbf{I}_{N^2}
\end{bmatrix}
\begin{bmatrix}
 \pmb{\xi} \\
 \pmb{\chi}
\end{bmatrix} \, .
\end{align}
Here, we define 
\begin{align}
\mathbf{S} =& \mathbf{I}_N \otimes D{f}(\mathbf{s}) + \sigma_1h(0)(r^{(1)}\mathbf{I}_N \otimes D {g}^{(1)})
 \notag\\
 &- 2\sigma_2( r^{(2)}\mathbf{I}_N \otimes D {g}^{(2)}) - \sigma_1h(0)\mathbf{L}^{(1)} \otimes D_{2} {g}^{(1)}\notag\\
&+ \sigma_2\mathbf{L}^{(2)} \otimes D_{s} {g}^{(2)}\, , 
\end{align}
with $D {g}^{(1)}=D_1{g}^{(1)}+ D_2{g}^{(1)}$, $D {g}^{(2)}=D_1{g}^{(2)}+ D_2{g}^{(2)} + D_3{g}^{(2)}$ and $D_{s} {g}^{(2)}= D_2{g}^{(2)} + D_3{g}^{(2)}$. $\mathbf{I}_N$ is the $N\times N$ identity matrix, $\mathbf{B}^{(1)}$ and $\mathbf{C}^{(1)}$ are suitable constant matrices of order $N \times N^2$ and $N^2 \times N$ (see Appendix \ref{sec:msf_calculation} for the explicit form of the latter) that satisfy $\mathbf{B}^{(1)}\mathbf{B}^{(1)^{\top}}=r^{(1)}\mathbf{I}_N$ and $\mathbf{B}^{(1)}\mathbf{C}^{(1)}=\mathbf{L}^{(1)}$. 

Solving the $(N^2+Nd)$ dimensional variational equation \eqref{pairvariational} to calculate the Lyapunov exponents provides the necessary condition for the stability of the synchronous solution. The high dimensionality of the variational equation prevents from an analytical study of the latter; however, the structure of the Jacobian matrix in Eq.~\eqref{pairvariational} indicates that there are $(N^2-N)$ eigenvalues equal to $-\epsilon$, hence $(N^2-N)$ stable directions with negative Lyapunov exponents, remember that $\epsilon>0$. The invariant subspace spanned by these eigenvalues allows us to introduce new coordinates, eventually reducing the dimension of \eqref{pairvariational} by separating the $(N^2-N)$ stable directions from the remaining $(Nd+N)$ ones. Therefore, the condition for the synchronization can be proved by studying the remaining $(Nd+N)$ dimensional system. Thereafter, in order to further simplify it and align with the strategy of the Master Stability Equation, we introduce the matrix $\mathbf{U}$ whose columns are the orthonormal eigenvectors of the Laplacian $\mathbf{L}^{(1)}$, i.e., $\mathbf{U}^{\top}\mathbf{L}^{(1)}\mathbf{U}=\mathrm{diag} \{0= \mu^{(1)}_{1}, \mu^{(1)}_{2}, \dots, \mu^{(1)}_{N} \}$. Then, we can perform the change of variables
\begin{align}
 \begin{pmatrix}
 \mathbf{U} \otimes \mathbf{I}_{d} & 0 \\
 0 & \mathbf{U}
 \end{pmatrix} \begin{pmatrix}
 \pmb{\xi} \\
 \pmb{\chi}
 \end{pmatrix}= \begin{pmatrix}
 \pmb{\zeta} \\
 \pmb{\eta}
 \end{pmatrix}\, , 
\notag\end{align}
and obtain, after some straightforward but lengthy computation that we postpone to the Appendix \ref{sec:msf_calculation}, the following master stability equation (MSE):
\begin{align}
\dot \zeta_{i} & = \Big[D{f}(\mathbf{s}) + \sigma_1 h(0)r^{(1)} D {g}^{(1)} - \sigma_1h(0) \mu^{(1)}_{i} D {g}^{(1)} + \notag \\ 
& - 2\sigma_2 r^{(2)} D{g}^{(2)}\Big]\zeta_{i} +\sigma_{2} \sum_{j=1}^{N} \tilde{L}^{2}_{ij}D_{s} {g}^{(2)}\zeta_{j}-\sigma_1{g}^{(1)}(\mathbf{s}, \mathbf{s})\eta_{i} \notag \\
\dot \eta_{i} & = -\epsilon \Big[ \mu^{(1)}_{i} Dh(0) \zeta_{i} + \eta_{i} \Big]\, , i=1, 2, \dots, N, 
\label{genmse1}
\end{align}
where we introduced the matrix $\tilde{\mathbf{L}}^{(2)}=\mathbf{U}^{\top}\mathbf{L}^{(2)}\mathbf{U}$. Therefore, the stability problem of the synchronous solution $(\mathbf{s}, k_{ij}^{(s)})$ is reduced to evaluating the maximum Lyapunov exponent of the above (non-autonomous) linear system of ODE~\eqref{genmse1}; let us recall that negativity of the latter implies the existence of a stable synchronous solution. Let us also observe that the MSE is still an $(Nd+N)$ dimensional coupled equation, and unlike the classical master stability approach, it cannot be further decoupled in $N$ equations of dimension $(d+1)$. Still, analogous to the classical master stability approach, we can separate the modes associated to the parallel and the transverse directions. The variables $(\zeta_1, \eta_1)$ correspond to the parallel modes, whereas the variables associated to $i=2, 3, \dots, N$ represent the transverse modes. Here, we use the fact that $\mu_1^{(1)}=0$ and $\mathbf{L}^{(2)}$ being a zero row sum matrix, the elements in the first row and column of the matrix $\tilde{\mathbf{L}}^{(2)}$ are zero. 

In general, the MSE can not be decoupled any further; there are, however, relevant cases in which one can take some steps further in the analytical understanding of the problem, as we will hereby show. In these scenarios, as in the classical master stability approach, we can decouple the MSE into 
$N$ equations of dimension $(d+1)$. 

\subsubsection{all-to-all connection topology}
We first consider the case where the higher-order Laplacian $\mathbf{L}^{(2)}$ is a scalar multiple of the pairwise Laplacian $\mathbf{L}^{(1)}$, i.e., $\mathbf{L}^{(2)}=\nu \mathbf{L}^{(1)}$ for some $\nu >0$. A straightforward example of this scenario occurs when the oscillators are globally connected to each other. For all-to-all connection involving $N$ nodes, we have $\mathbf{L}^{(2)} = (N-1)\mathbf{L}^{(1)}$. Therefore, in this case Eq.~\eqref{genmse1} can be simplified and returns for all $i=1, 2, \dots, N$
\begin{align}
 \dot \zeta_{i} & = [D{f}(\mathbf{s}) + \sigma_1h(0)r^{(1)} D {g}^{(1)} - \sigma_1h(0) \mu^{(1)}_{i} D {g}^{(1)}+ \notag \\ 
 - & 2\sigma_2 r^{(2)} D{g}^{(2)} +\nu \mu^{(1)}_{i}D_{s} {g}^{(2)}]\zeta_{i}-\sigma_1{g}^{(1)}(\mathbf{s}, \mathbf{s})\eta_{i}, \notag\\
 \dot \eta_{i} & = -\epsilon[ \mu^{(1)}_{i} Dh(0) \zeta_{i} + \eta_{i} ]\,.
 \label{mseglobal}
\end{align}
Since, for the globally coupled topology, $\nu=(N-1)$ and $\mu^{(1)}_i=N$, for all $i=2, 3, \dots, N$, to find the stability of the system, we just need to find the maximum Lyapunov exponent ($\lambda_{max}$) of the $(d+1)$ dimensional equation
\begin{align}
 \dot \zeta & = [D{f}(\mathbf{s}) + \sigma_1h(0)r^{(1)} D {g}^{(1)} - \sigma_1h(0) (N-1) D {g}^{(1)}+ \notag \\ 
 - & 2\sigma_2 r^{(2)} D{g}^{(2)} +N(N-1) D_{s} {g}^{(2)}]\zeta-\sigma_1{g}^{(1)}(\mathbf{s}, \mathbf{s})\eta, \notag\\
 \dot \eta & = -\epsilon[ N Dh(0) \zeta + \eta ]\ .
 \label{mseglobal2}
\end{align}

\subsubsection{Commuting Laplacian matrices}
Another interesting case where one can simplify the MSE, is the one where the pairwise and higher-order Laplacians commute each other, i.e., $\mathbf{L}^{(1)}\mathbf{L}^{(2)}=\mathbf{L}^{(2)}\mathbf{L}^{(1)}$. Let us observe that such a scenario can be obtained by assuming the underlying supports to have nonlocal ring-like topology, where each node is connected to $k$ neighboring nodes on the left and $k$ on the right via pairwise or higher-order connections. Under this assumption the matrices $\mathbf{L}^{(1)}$ and $\mathbf{L}^{(2)}$ are circulant matrices and thus they commute. The latter conclusion implies that the matrix $\mathbf{U}$ diagonalizes also $\mathbf{L}^{(2)}$, i.e., $\mathbf{U}^{\top}\mathbf{L}^{(2)}\mathbf{U}=\mathrm{diag} \{\mu^{(2)}_{1}, \mu^{(2)}_{2}, \dots, \mu^{(2)}_{N} \}$. Eventually the MSE~\eqref{genmse1} simplifies into
\begin{align}
 \dot \zeta_{i} & = [D{f}(\mathbf{s}) + \sigma_1h(0)r^{(1)} D {g}^{(1)} - \sigma_1h(0) \mu^{(1)}_{i} D_{s} {g}^{(1)}+ \notag\\ 
 - & 2\sigma_2 r^{(2)} D{g}^{(2)} +\mu^{(2)}_{i}D_{s} {g}^{(2)}]\zeta_{i}-\sigma_1{g}^{(1)}(\mathbf{s}, \mathbf{s})\eta_{i}, \label{msenonlocala}\\
 \dot \eta_{i} & = -\epsilon[ \mu^{(1)}_{i} Dh(0) \zeta_{i} + \eta_{i} ]\, , i=1, 2, \dots, N.
 \label{msenonlocalb}
\end{align}
Thus, the MSE decouples into $N$ blocks of $(d+1)$-dimensional equations. Solving these low dimensional equations for the calculation of maximum Lyapunov exponent, $\lambda_{max}$, provides the necessary conditions for the stability of the synchronous solution. It is important to note that $\lambda_{max}$ depends only on the coupling strengths and the structural properties of the connection topology, via the eigenvalues and the (generalized) degrees, resembling thus to the classical MSE approach. 

\subsection{Numerical Results}
Let us now present some numerical results to support the theory presented above. Because we will not make any assumption on the coupling function, in particular, they will not be noninvasive, we recall that the underlying support should satisfy $\sum_{j=1}^{N} a_{ij}^{(1)}=r^{(1)}$ and $\sum\limits_{j=1}^{N}\sum\limits_{p=1}^{N} a_{ijp}^{(2)}=2r^{(2)}$, for all $i$, for the synchronous solution to exist. To perform the numerical simulations, we integrate the adaptive higher-order system using the fourth-order Runge-Kutta (RK4) algorithm with adaptive time-stepping, maintaining a relative tolerance of $10^{-8}$ for $T=10^{4}$ time units and unless stated otherwise, the number of oscillators is set fixed to $N=200$. 

\subsubsection{Kuramoto oscillators with all-to-all topology} 

The first example we propose is a Kuramoto model with pairwise and three-body interactions and phase lags \cite{berner2021desynchronization, berner2019multiclusters} defined via an all-to-all coupling. %
\begin{figure}[!ht]
\centering\includegraphics[width=0.99\linewidth]{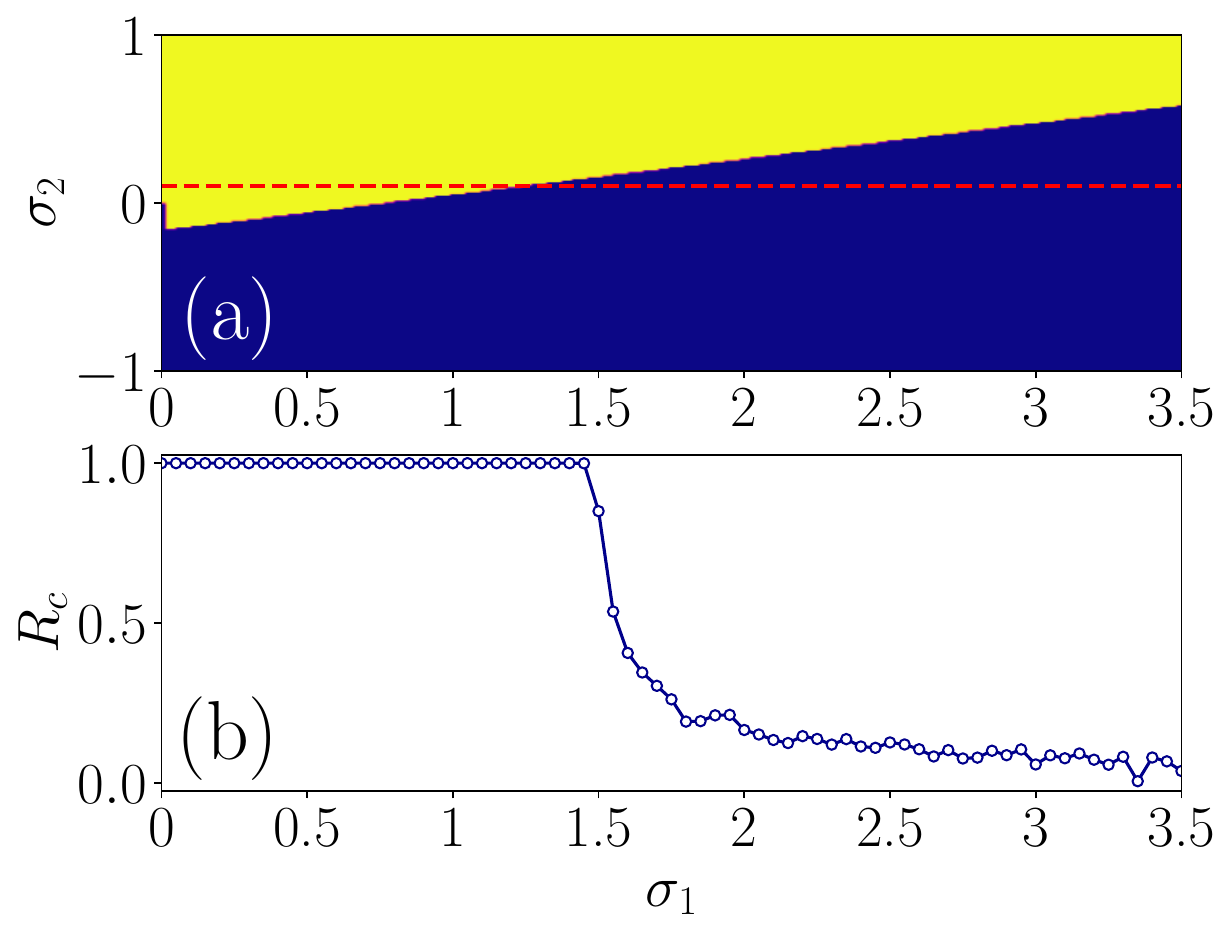}
\caption{
(a) Phase diagram in the $\sigma_1-\sigma_2$ plane illustrating synchronization in the Kuramoto model with lag and all-to-all coupling (model I). Regions associated with positive Lyapunov exponents are colored blue (dark), while yellow (bright) zones indicate negative MSF. (b) The order parameter $R_{c}$ is plotted as a function of $\sigma_{1}$ for $\sigma_{2}=0.1$, starting from the synchronization state with initial conditions close to $\theta_{i}(0)=0$ and $k_{ij}^{(1)}=-a_{ij}^{(1)}\sin(\beta_{1})$ indicated by the horizontal red-dashed line in (a). It is observed that the MSF becomes positive precisely at the value of $\sigma_{1}$ where $R_{c}<1$, signifying a decrease from synchrony. Other parameters are $N=200$, $\alpha_{1} = \alpha_{2} = 0.49\pi$, $\beta_{1}=0.88\pi$, $\epsilon_1=0.01$. The order parameter$R_{c}$ is calculated by averaging the last $25\%$ of the datas.
} 
 \label{fig:figKlagnohoi}
\end{figure}
The system's state is thus described by $N$ angular variables, $\theta_i$, and $N^2$ links weights; we recall that we are still assuming the three-body interaction to not evolve and thus be fixed to $1$. In equations, 
\begin{align}
\dot \theta_{i} =& \; \omega - \frac{\sigma_1}{N}\sum_{j=1}^N a_{ij}^{(1)} k_{ij}^{(1)} \sin(\theta_i - \theta_j + \alpha_1)+ \notag \\ 
& - \frac{\sigma_2}{N^2} \sum_{j=1}^N \sum_{p=1}^N a_{ijp}^{(2)} \sin(2 \theta_i - \theta_j - \theta_p + \alpha_2), \label{pairkuratheta}\\
\dot k_{ij}^{(1)} =& -\epsilon_{1} [k_{ij}^{(1)} + a_{ij}^{(1)} \sin(\theta_i - \theta_j + \beta_1)] \label{pairkurakij}\, , 
\end{align}
where $\alpha_1>0$ (resp. $\alpha_2$) determines the pairwise (resp. three-body) lag, while $\beta_1$ is the lag in the evolution of links weights. To allow for a global synchronous solution we assume the proper frequency to be equal each other and without any loss of generality, we consider $\omega=0$ here and throughout the rest of the manuscript by choosing a suitable frame of reference.

The global synchronous solution is given by
\begin{align}
\mathbf{s}(t)=& \bigg(\frac{\sigma_1}{N}r^{(1)}\sin\alpha_1\sin\beta_1 - \frac{\sigma_2}{N^2}2r^{(2)}\sin\alpha_2 \bigg)t, \\
 k_{ij}^{(s)} =& - a_{ij}^{(1)} \sin\beta_1\, , 
\end{align}
whose stability can be studied by using Eq.~\eqref{mseglobal2}, namely 
\begin{align}
\begin{bmatrix}
 \dot{\mathbf{{\zeta}}} \\
 \dot{\eta}
\end{bmatrix} 
= 
\begin{bmatrix}
 M_{11} & M_{12}\\
 M_{21} & M_{22}
\end{bmatrix}
\begin{bmatrix}
 \mathbf \zeta \\
 \eta
\end{bmatrix}\, ,
\end{align}
where 
\begin{eqnarray*}
M_{11}&=&\sigma_1\cos\alpha_1\sin\beta_1- 2\frac{\sigma_2}{N}(N-1)\cos\alpha_2\\ M_{12}&=&-\frac{\sigma_1}{N}\sin\alpha_1 \, ,
 M_{21}= -\epsilon_{1} N\cos\beta_1 \text{ and } M_{22}=-\epsilon_{1} \, .
\end{eqnarray*}

The characteristic polynomial of the latter system is
\begin{align}
 \lambda^2 + & \lambda\left(\epsilon_{1} - \sigma_1\cos\alpha_1\sin\beta_1 + 2\frac{\sigma_2}{N}(N-1)\cos\alpha_2\right) \notag \\ - &\epsilon_{1} \left(\sigma_1\sin(\alpha_1 + \beta_1) - 2\frac{\sigma_2}{N}(N-1)\cos\alpha_2\right) = 0\, , 
\end{align}
whose roots can be straightforwardly computed, and thus, the stability of the synchronous solution can be inferred. The results are reported in Fig.~\ref{fig:figKlagnohoi} (a), where we show the synchronization region as a function of the coupling parameters $\sigma_1$ and $\sigma_2$ by using a color code: yellow $\max(\Re\lambda) <0$ and blue otherwise. It can be observed that for $\sigma_{2} \geq 0$, the system goes through a transition from synchronization to desynchronization until a critical value of $\sigma_{2} \approx 0.58$. Beyond this, the system always remains in synchrony. On the other hand, for $\sigma_{2}<0$, up to a small range of $\sigma_{2}$ ($\approx -0.15$), a bounded region of synchronization can be observed, i.e., at first a transition from desynchrony to synchrony is observed within a small window of $\sigma_{1}$, and then synchrony to desynchrony emerges. Beyond this, the system never settles in a stable synchronous solution. 

To perform numerical investigations, we utilize the cluster order parameter $R_{c}$ \cite{Kasatkin2017, Kasatkin2018, berner2021desynchronization} to measure the system coherence. $R_{c}$ is given by the number of pairwise coherent oscillators normalized by the total number of pairs $N^2$. Mathematically, it is defined as, 
\begin{align}
\label{eq:Rc}
R_c = \frac{1}{N^2}\sum_{i, j=1}^NR_{ij}\, , 
\end{align}
where
\begin{align}
R_{ij} = \left\vert\frac{1}{\Delta t}\int_{T-\Delta t}^{T}\exp{\mathrm{i} [\theta_i(t) - \theta_j(t)]}dt\right\vert\, .
\end{align}
Thus, in an incoherent (resp. fully coherent) state, $R_{c}$ returns the values $0$ (reps. $1$). While for the frequency cluster states, it takes the value $0<R_{c}<1$. 
Figure~\ref{fig:figKlagnohoi}(b) portrays the variation of $R_{c}$ as a function of $\sigma_{1}$ for a fixed higher-order coupling, $\sigma_{2}=0.1$; the system has been intialized  starting from the synchronization state with initial conditions close to $\theta_{i}(0)=0$ and $k_{ij}^{(1)}=-a_{ij}^{(1)}\sin(\beta_{1})$. 
We can observe the existence of a transition from full synchronization to desynchronization through intermediate cluster states (decreasing $R_c$), which shows good agreement with the analytical findings. 

Let us now consider a slightly different higher-order Kuramoto model still involving pairwise and three-body interactions defined through all-to-all coupling topology. %
\begin{figure}[!ht]
\centering\includegraphics[width=0.99\linewidth]{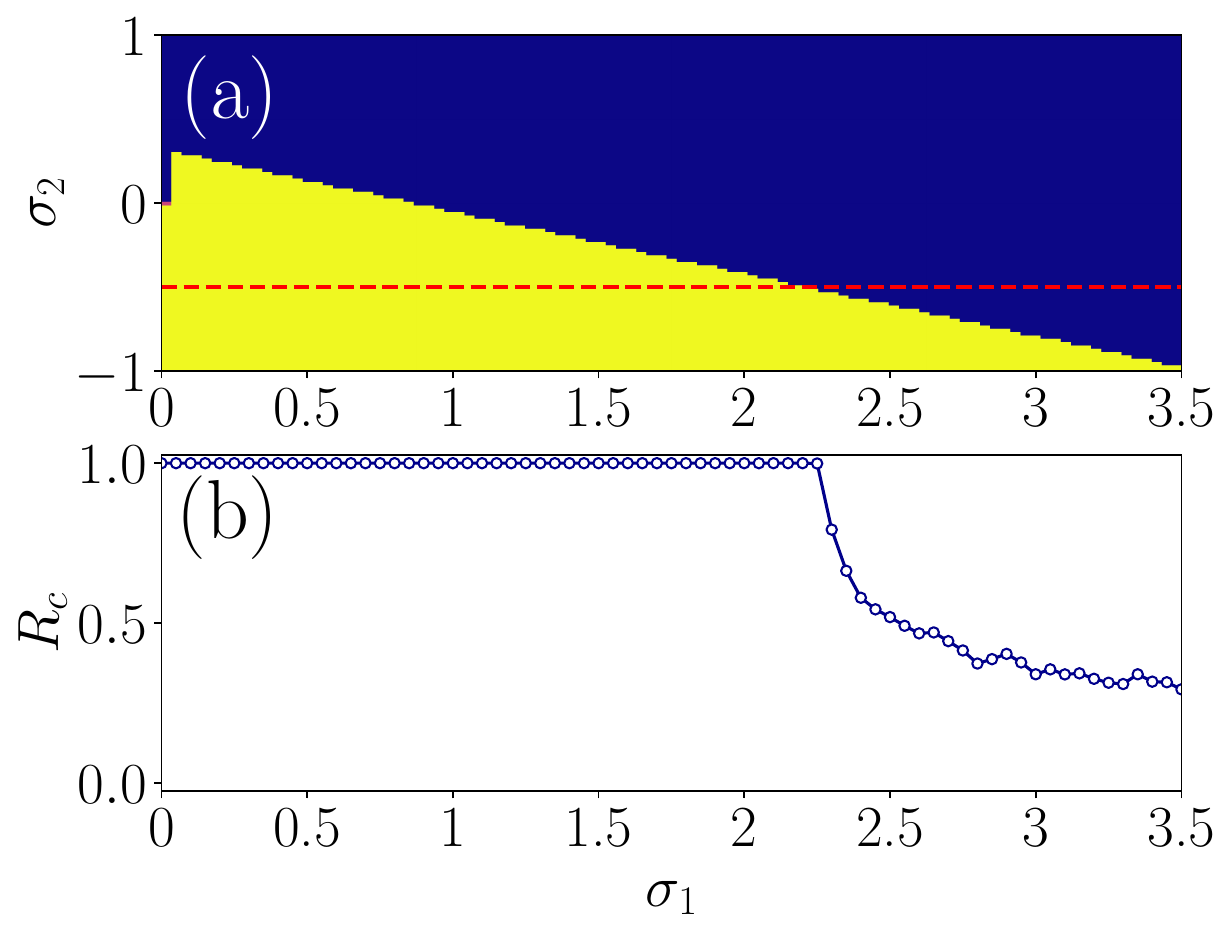}
\caption{
(a) Phase diagram of the MSF in the $\sigma_1-\sigma_2$ plane (coupling parameters), illustrating synchronization behavior in the Kuramoto model with lag and all-to-all coupling (Model II).  In the diagram, regions with positive Lyapunov exponents are shown in blue (dark), indicating instability, while yellow (bright) zones represent regions of negative MSF, corresponding to stability. (b) The order parameter  $R_{c}$ is shown as a function of $\sigma_{1}$ for $\sigma_{2}=-0.5$ starting from the synchronization state with initial conditions close to $\theta_{i}(0)=0$ and $k_{ij}^{(1)}=-a_{ij}^{(1)}\sin(\beta_{1})$ corresponding to the horizontal red-dashed line in (a). The MSF is observed to become positive precisely at the $\sigma_1$ value where $R_{c}<1$, signifying a decrease from synchrony. Other model parameters are kept the same as in Fig.~\ref{fig:figKlagnohoi}
}
 \label{fig:figKlagnohoiII}
\end{figure}
In this case, the chosen higher-order interaction term has an asymmetric form, similar to those recently used by several researchers \cite{skardal2020higher}. The dynamics of the system is given by
\begin{align}
\dot \theta_{i} =& \omega - \frac{\sigma_1}{N}\sum_{j=1}^N a_{ij}^{(1)} k_{ij}^{(1)} \sin(\theta_i - \theta_j + \alpha_1)+ \notag \\ 
 -& \frac{\sigma_2}{N^2} \sum_{j=1}^N \sum_{p=1}^N a_{ijp}^{(2)} \sin(2 \theta_j - \theta_p - \theta_i + \alpha_2), \\
\dot k_{ij}^{(1)} =& -\epsilon_{1} [k_{ij}^{(1)} + a_{ij}^{(1)} \sin(\theta_i - \theta_j + \beta_1)] \, .
\end{align}
The synchronous solution is
\begin{align}
\mathbf{s}(t) =& \bigg(\frac{\sigma_1}{N}r^{(1)}\sin\alpha_1\sin\beta_1 - \frac{\sigma_2}{N^2}2r^{(2)}\sin\alpha_2 \bigg)t, \\
 k_{ij}^{(s)} =& - a_{ij}^{(1)} \sin\beta_1\, , 
\end{align}
whose stability can be studied by applying strategy similar to the one used above and thus returning the linear system
\begin{align}
\begin{bmatrix}
 \dot{\mathbf{{\zeta}}} \\
 \dot{\eta}
\end{bmatrix} 
= 
\begin{bmatrix}
 M_{11} & M_{12}\\
 M_{21} & M_{22}
\end{bmatrix}
\begin{bmatrix}
 \mathbf \zeta \\
 \eta
\end{bmatrix}\, ,
\end{align}
where 
\begin{eqnarray*}
    M_{11}&=&\sigma_1\cos\alpha_1\sin\beta_1 + \frac{\sigma_2}{N}(N-1)\cos\alpha_2, \\ M_{12}&=&-\frac{\sigma_1}{N}\sin\alpha_1, M_{21}= -\epsilon_{1} N \cos\beta_1 \;\text{and} \;M_{22}=-\epsilon_{1}.
\end{eqnarray*}
The corresponding characteristics polynomial to the above linear system is
\begin{align}
 \lambda^2 + \left(\epsilon_{1} - \sigma_1\cos\alpha_1\sin\beta_1 - \frac{\sigma_2}{N}(N-1)\cos\alpha_2\right)\lambda \notag \\ -\epsilon_{1} \left(\sigma_1\sin(\alpha_1 + \beta_1) + \frac{\sigma_2}{N}(N-1)\cos\alpha_2\right) = 0\, .
\end{align}
In Fig.~\ref{fig:figKlagnohoiII} we report the MSF \textcolor{blue}{[cf. Fig.~\ref{fig:figKlagnohoiII}(b)]} computed again as a function of $\sigma_1$ and $\sigma_2$ and compared with the order parameter $R_c$ \textcolor{blue}{[cf. Fig.~\ref{fig:figKlagnohoiII}(a)]}. Here, we can observe a relatively opposite scenario as compared to the results of the first model. It can be remarked that for $\sigma_{2} \leq 0$, the system goes through a transition from synchronization to desynchronization until a critical value of $\sigma_{2} \approx -0.97$ is reached. Beyond this value, the system always remains synchronized. On the other hand, for $\sigma_{2}>0$, up to a small range of $\sigma_{2} (\approx 0.3)$, a bounded region of synchronization can be observed, beyond which the system never settles in a stable synchronous solution. 

\subsubsection{Kuramoto oscillators with nonlocal connection topology} 
To move forward, we again consider the higher-order Kuramoto model, given by Eqs.~\eqref{pairkuratheta} and \eqref{pairkurakij} but with the pairwise and higher-order coupling realized via a nonlocal connection topology. More precisely, the oscillators are connected through a ring-like topology where each node is connected to $k=30$ neighboring nodes on the left and $k=30$ on the right via pairwise connections. The associated triadic connections are formed by promoting all the triangles composed of three different pairwise links. In this way, the oscillators can engage in both pairwise interactions, when connected by a network link, and three-body interactions, when they are part of the same triangle. By following Eqs.~\eqref{msenonlocala} and \eqref{msenonlocalb}, the stability of the synchronous solution can be inferred by solving the roots of the characteristics equation corresponding to each one of the $(N-1)$ blocks (since the first block is associated with the parallel modes). The characteristics equation for each block $i=2, 3, \dots, N$ is given by, 
\begin{align}
 \lambda_i^2 + & \lambda_i\left(\epsilon_{1} - \frac{\sigma_1}{N}\mu_{i}^{(1)}\cos\alpha_1\sin\beta_1 + 2\frac{\sigma_2}{N^2}\mu_{i}^{(2)}\cos\alpha_2\right) \notag \\ - &\epsilon_{1}\mu_{i}^{(1)} \left(\frac{\sigma_1}{N}\sin(\alpha_1 + \beta_1) - 2\frac{\sigma_2}{N^2}(N-1)\cos\alpha_2\right) = 0\, .
\end{align}

\begin{figure}[!ht]
\centering\includegraphics[width=0.99\linewidth]{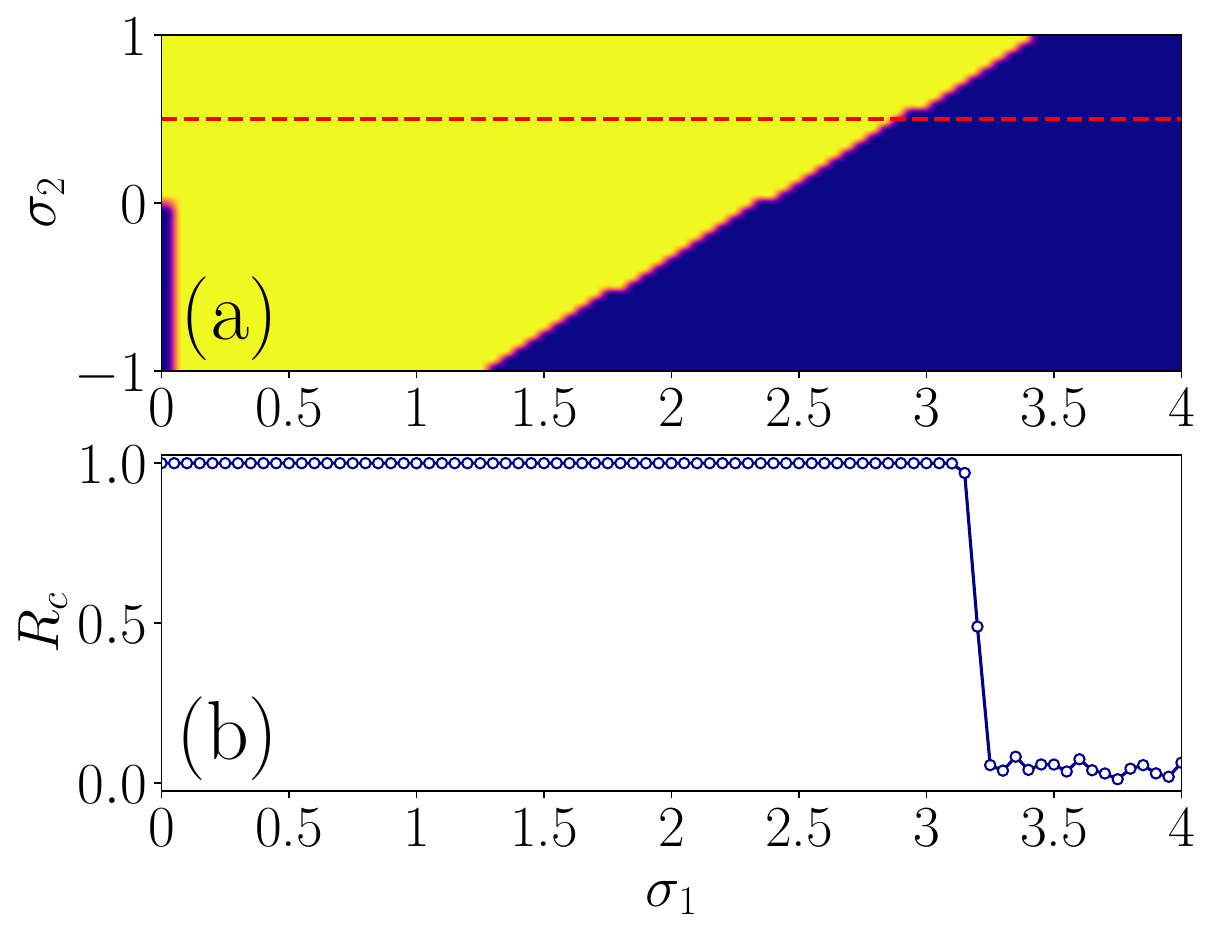}
\caption{
(a) Phase diagram of the MSF in the $\sigma_1-\sigma_2$ plane (coupling parameters) illustrating synchronization in the Kuramoto model with lag and nonlocal coupling topology. Regions with positive Lyapunov exponents are colored blue (dark), while yellow (bright) zones indicate negative MSF. (b) Plot of the order parameter $R_{c}$ as a function of $\sigma_{1}$ for $\sigma_{2}=0.5$ starting from the synchronization state with initial conditions close to $\theta_{i}(0)=0$ and $k_{ij}^{(1)}=-a_{ij}^{(1)}\sin(\beta_{1})$, as represented by the horizontal red-dashed line in (a). The remaining model parameters are kept constant as in Fig.~\ref{fig:figKlagnohoi}, with the number of connected oscillators on both sides set to $k=30$
} 
\label{fig:figKlagnohoinonlocal}
\end{figure}

A negative value of $\lambda_{max}=\text{max} \{\Re \lambda_{i} \} $ signifies the emergence of stable synchronous solution. Results are reported in Fig.~\ref{fig:figKlagnohoinonlocal} where  $\lambda_{max}$ (panel (a)) is computed as a function of $\sigma_1$ and $\sigma_2$ and compared with the order parameter $R_c$ (panel (b)) for a fixed $\sigma_{2}$. It can be observed that the system goes through a transition from full synchronization to desynchronization with increasing $\sigma_{1}$ for $\sigma_{2}\geq 0$. On the other hand, for $\sigma_{2}<0$, the system achieves synchronization in a bounded region. It is also worth noticing that with increasing $\sigma_{2}$ from negative to positive, the synchronization region becomes much wider. 

\section{Adaptive pairwise and higher-order interactions}
\label{sec:adaptprwshoi}
The aim of this section is to generalize the previous study by relaxing the assumption about the static higher-order interactions; more precisely, we now assume the pairwise weights and the higher-order ones to evolve as functions of the system state. More precisely, by restricting again, for the sake of pedagogy, our analysis to the three-body interaction, we are considering the model
\begin{align}
\dot{\mathbf{x}}_i =& {f}(\mathbf{x}_i) - \sigma_1 \sum_{j=1}^N a_{ij}^{(1)} k_{ij}^{(1)}(t) {g}^{(1)}(\mathbf{x}_i, \mathbf{x}_j) +\notag \\ & -\sigma_2{\sum_{j=1}^N}{\sum_{p=1}^N} a_{ijp}^{(2)} k_{ijp}^{(2)}(t){g}^{(2)}(\mathbf{x}_i, \mathbf{x}_j, \mathbf{x}_p)\, , 
\end{align}
where ${g}^{(1)}(\mathbf{x}_i, \mathbf{x}_j)$ and ${g}^{(2)}(\mathbf{x}_i, \mathbf{x}_j, \mathbf{x}_k)$ are the coupling functions, and the weights of pairwise, $k_{ij}^{(1)}$, and three-body interactions, $k_{ijp}^{(2)}$, evolve as follows
\begin{align}
\label{eq:2}
\dot k_{ij}^{(1)} = & -\epsilon_1[k_{ij}^{(1)} + a_{ij}^{(1)} {h}^{(1)}(\mathbf{x}_i-\mathbf{x}_j)], \\
\dot k_{ijp}^{(2)} = &-\epsilon_2[k_{ijp}^{(2)} + a_{ijp}^{(2)} {h}^{(2)}(2\mathbf{x}_i- \mathbf{x}_j- \mathbf{x}_p)]\, .
\end{align}
Here, ${h}^{(1)}$ and ${h}^{(2)}$ denote adaptation functions. $\epsilon_{i}$,  $i=1, 2$ is the time scale separation parameter, which is generally taken to be a very small positive real number.

The global synchronous solution of the above adaptive model is given by $(\mathbf{s}, k_{ij}^{(s)}, k_{ijk}^{(s)})$ and it satisfies
\begin{align}
\dot{\mathbf{s}} = & {f}(\mathbf{s}) + \sigma_1 r^{(1)}h^{(1)}(0) {g}^{(1)}(\mathbf{s}, \mathbf{s}) \notag \\
& + 2\sigma_2r^{(2)}h^{(2)}(0){g}^{(2)}(\mathbf{s}, \mathbf{s}, \mathbf{s}), \\ 
k_{ij}^{(s)} =& -a_{ij}^{(1)}{h}^{(1)}(0), \\ 
k_{ijp}^{(s)} =& -a_{ijp}^{(2)}{h}^{(2)}(0) \, .
\end{align}
It is important to note that here, once again, we have not imposed any conditions on the coupling functions, and thus, the underlying connectivity has constant (generalized) degrees $r^{(i)}, \; i=1, 2$. The analysis with non-invasive coupling functions is illustrated in Appendix~\ref{sec:noninvasive}. 

A linear stability analysis about this solution can be performed to infer the existence of a stable synchronous solution. We thus perturb the system with small perturbation terms $\mathbf{\xi}_i = \mathbf{x}_i - \mathbf{s} $, $ \chi_{ij} = k_{ij}^{(1)} - k_{ij}^{(s)} $ and $\eta_{ijp} = k_{ijp}^{(2)} - k_{ijp}^{(s)}$, and we study their time evolution via the variational equations,
\begin{widetext}
\begin{align}
\label{5:1}
\dot{\mathbf{\xi}}_i = & D{f}(\mathbf{s})\mathbf{\xi}_i +\sigma_{1}r^{(1)}h^{(1)}(0){D}{g}^{(1)}(\mathbf{s}, \mathbf{s}) \xi_i +2\sigma_{2}r^{(2)}h^{(2)}(0){D}{g}^{(2)}(\mathbf{s}, \mathbf{s}, \mathbf{s})\xi_{i} -\sigma_1 {h}^{(1)}(0)\sum_{j=1}^N {L}_{ij}^{(1)} {D}_2{g}^{(1)}(\mathbf{s}, \mathbf{s})\xi_j + \notag \\
& - \sigma_2{h}^{(2)}(0)\sum_{j=1}^N {L}_{ij}^{(2)}{D}_s{g}^{(2)}(\mathbf{s}, \mathbf{s}, \mathbf{s})\xi_j - \sigma_1 \sum_{j=1}^N a_{ij}^{(1)} \chi_{ij} {g}^{(1)}(\mathbf{s}, \mathbf{s}) - \sigma_2\sum_{j=1}^N \sum_{p=1}^N a_{ijp}^{(2)}\eta_{ijp}{g}^{(2)}(\mathbf{s}, \mathbf{s}, \mathbf{s}), \\
 \dot{\chi}_{ij} = & - \epsilon_1(\chi_{ij} + a_{ij}^{(1)}{Dh}^{(1)}(0)(\xi_i - \xi_j)), \\
 \dot{\eta}_{ijp} =& - \epsilon_2(\eta_{ijp} + a_{ijp}^{(2)}{Dh}^{(2)}(0)(2\xi_i - \xi_j - \xi_p))\, , 
\end{align}
which is a $(N^3+N^2+Nd)$-dimensional coupled linear differential equation. In matrix form, the above variational equations can be written as
\begin{align}
\begin{bmatrix}
 \dot{\pmb{\xi}} \\
 \dot{\pmb{\chi}} \\
 \dot{\pmb{\eta}}
\end{bmatrix} 
= 
\begin{bmatrix}
 \mathbf{S}_2 & -\sigma_1 \mathbf{B}^{(1)} \otimes {g}^{(1)}(\mathbf{s}, \mathbf{s}) & -\sigma_2 \mathbf{B}^{(2)} \otimes {g}^{(2)}(\mathbf{s}, \mathbf{s}, \mathbf{s}) \\
 -\epsilon_{1} \mathbf{C}^{(1)} \otimes Dh^{(1)}(0) & -\epsilon_{1} \mathbf{I}_{N^2} & 0 \\
 -\epsilon_{2} \mathbf{C}^{(2)} \otimes Dh^{(2)}(0) & 0 & -\epsilon_{2} \mathbf{I}_{N^3} 
\end{bmatrix} 
\begin{bmatrix}
\pmb{\xi} \\
\pmb{\chi} \\
\pmb{\eta}
\end{bmatrix} \, , 
\end{align}
\end{widetext}
where we consider again the matrix $\pmb{\chi}$, resp. the tensor $\pmb{\eta}$, as $N^2$, resp. $N^3$, columns vectors by stacking the rows over rows similarly to the previous case, and we introduced suitable matrices
\begin{align}
\mathbf{S}_2 = & \mathbf{I}_N \otimes D{f}(\mathbf{s}) + \sigma_1h^{(1)}(0)(r^{(1)}\mathbf{I}_N \otimes D {g}^{(1)}) \notag \\
& + 2\sigma_2h^{(1)}(0)( r^{(2)}\mathbf{I}_N \otimes D {g}^{(2)}) - \sigma_1h^{(1)}(0)\mathbf{L}^{(1)} \otimes D_{s} {g}^{(1)} \notag \\ 
& - \sigma_2h^{(2)}(0)\mathbf{L}^{(2)} \otimes D_{s} {g}^{(2)}, 
\end{align}
$\mathbf{B}^{(1)}$ and $\mathbf{C}^{(1)}$ are the same constant matrices of order $N \times N^2$ and $N^2 \times N$ used in Section~\ref{sec:nohoiadapt}, while $\mathbf{B}^{(2)}$ and $\mathbf{C}^{(2)}$ are constant matrices of order $N \times N^3$ and $N^3 \times N$, satisfying $\mathbf{B}^{(2)}\mathbf{B}^{(2)^{\top}}=2r^{(2)}\mathbf{I}_N$ and $\mathbf{B}^{(2)}\mathbf{C}^{(2)}=\mathbf{L}^{(2)}$ (we refer the interested reader to Appendix \ref{sec:msf_calculation} for a longer description of the derivation of those matrices).

To move forward and to be able to get some analytical insight, we hereafter assume $\epsilon_1=\epsilon_2$. Let us observe that one can relax this assumption by considering noninvasive coupling functions $g^{(1)}$ and $g^{(2)}$ (see Appendix \ref{sec:noninvasive}). Leaving the details of the computation to the Appendix \ref{sec:msf_calculation}, we can eventually obtain the $(Nd+2N)$-dimensional MSE ruling the evolution of the perturbations
\begin{align}
 \dot{\hat{\xi}}_{i}=& \left[D{f}(\mathbf{s}) + \sigma_1h^{(1)}(0)r^{(1)} D {g}^{(1)} \right. \notag\\ 
 & \left. + 2\sigma_2 r^{(2)}h^{(2)}(0) D{g}^{(2)} + \sigma_1h^{(1)}(0) \mu^{(1)}_{i} D_{s} {g}^{(1)} \right]\hat{\xi}_{i} \notag \\ 
 & -\sigma_{2}h^{(2)}(0)\sum\limits_{j=1}^{N} \Tilde{L}^{2}_{ij}D_{s} {g}^{(2)}\hat{\xi}_{j} -\sigma_1{g}^{(1)}(s, s)\hat{\chi}_{i} \notag \\ 
 & -\sigma_2{g}^{(2)}(s, s, s)\hat{\eta}_{i}, \notag \\
 \dot{\hat{\chi}}_{i} =& -\epsilon[ \mu^{(1)}_{i} Dh^{(1)}(0)\hat{\xi}_{i} + \hat{\chi}_{i} ], \notag \\
 \dot{\hat{\eta}}_{i}=& -\epsilon[ \sum\limits_{j=1}^{N} \Tilde{L}^{2}_{ij} Dh^{(2)}(0)\hat{\xi}_{i} + \hat{\eta}_{i} ]\, , 
 \label{genmse2}
\end{align}
where we introduced new coordinates $(\hat{\pmb{\xi}}, \hat{\pmb{\chi}}, \hat{\pmb{\eta}})^\top$ related to the previous ones by
\begin{align}
 \begin{pmatrix}
 \mathbf{U} \otimes \mathbf{I}_{d} & 0 & 0\\
 0 & \mathbf{U} & 0 \\
 0 & 0 & \mathbf{U}
 \end{pmatrix} \begin{pmatrix}
 \pmb{\xi} \\
 \pmb{\chi} \\
 \pmb{\eta}
 \end{pmatrix}= \begin{pmatrix}
 \hat{\pmb{\xi}} \\
 \hat{\pmb{\chi}} \\
 \hat{\pmb{\eta}}
 \end{pmatrix}, 
\notag\end{align}
with $\mathbf{U}$ the matrix whose columns are the orthonormal eigenvectors that diagonalizes the pairwise Laplacian $\mathbf{L}^{(1)}$, i.e., $\mathbf{U}^{\top}\mathbf{L}^{(1)}\mathbf{U}=\mathrm{diag} \{\mu^{(1)}_{1}, \mu^{(1)}_{2}, \dots, \mu^{(1)}_{N} \}$ and $\tilde{\mathbf{L}}^{(2)}=\mathbf{U}^{\top}\mathbf{L}^{(1)}\mathbf{U}$. 

Therefore, the stability problem of the synchronous solution is reduced to evaluating the maximum Lyapunov exponent of the above coupled linear differential equation \eqref{genmse2}. Once again, we are faced with the problem that, in general, the MSE~\eqref{genmse2} can not be decoupled further. Still, analogous to the classical master stability approach \cite{fujisaka1983stability,msf}, we can separate the modes associated with parallel and transverse directions. The variables $(\hat{\xi}_{1}, \hat{\chi}_{1}, 
\hat{\eta}_{1})$ correspond to the parallel models, whereas the variables associated with $i=2, 3, \dots, N$ represent the transverse modes. Here, we once again use the fact that $\mu_1^{(1)}=0$ and $\mathbf{L}^{(2)}$ being a zero row sum matrix, the elements in the first row and column of the matrix $\tilde{\mathbf{L}}^{(2)}$ are zero. In general, the transverse modes of the MSE~\eqref{genmse2} can not be further separated. However, there are interesting cases in which the MSE can be fully decoupled, similar to what has been addressed for the only pairwise adaptation case, i.e., when the higher-order Laplace matrix is a scalar multiple of the pairwise one and the scenario of commuting Laplacian matrices. In these cases, thus, the MSE can be separated into $N$ $(d+2)$-dimensional equations, and the maximum Lyapunov exponents depend only on the interaction coupling strengths and the underlying structure properties via the eigenvalues of the Laplace matrices and the (generalized) degrees. 

For the all-to-all coupling configuration similar to the only pairwise adaptation case, the MSE reduced into $(d+2)$-dimensional equations given by
\begin{align}
 \dot{\hat{\xi}}=& \left[D{f}(\mathbf{s}) + \sigma_1h^{(1)}(0)r^{(1)} D {g}^{(1)} \right. \notag\\ 
 & \left. + 2\sigma_2 r^{(2)}h^{(2)}(0) D{g}^{(2)} + \sigma_1h^{(1)}(0) \mu^{(1)} D_{s} {g}^{(1)} \right]\hat{\xi}_{i} \notag \\ 
 & -\sigma_{2}h^{(2)}(0)\nu \mu^{(1)}D_{s} {g}^{(2)}\hat{\xi}_{i} -\sigma_1{g}^{(1)}(s, s)\hat{\chi} \notag \\ 
 & -\sigma_2{g}^{(2)}(s, s, s)\hat{\eta}, \notag \\
 \dot{\hat{\chi}} =& -\epsilon[ \mu^{(1)} Dh^{(1)}(0)\hat{\xi} + \hat{\chi} ], \notag \\
 \dot{\hat{\eta}}=& -\epsilon[ \nu \mu^{(1)} Dh^{(2)}(0)\hat{\xi} + \hat{\eta} ]\, , 
 \label{genmseglobl}
\end{align}
where $\nu=N-1$ and $\mu^{(1)}=N$. Thus, the problem of stability analysis is reduced to solve the $(d+2)$-dimensional equation for the calculation of the maximum Lyapunov exponent.

\subsection{Numerical Analysis}
\begin{figure*}[htp]
\centering\includegraphics[width=0.88\linewidth]{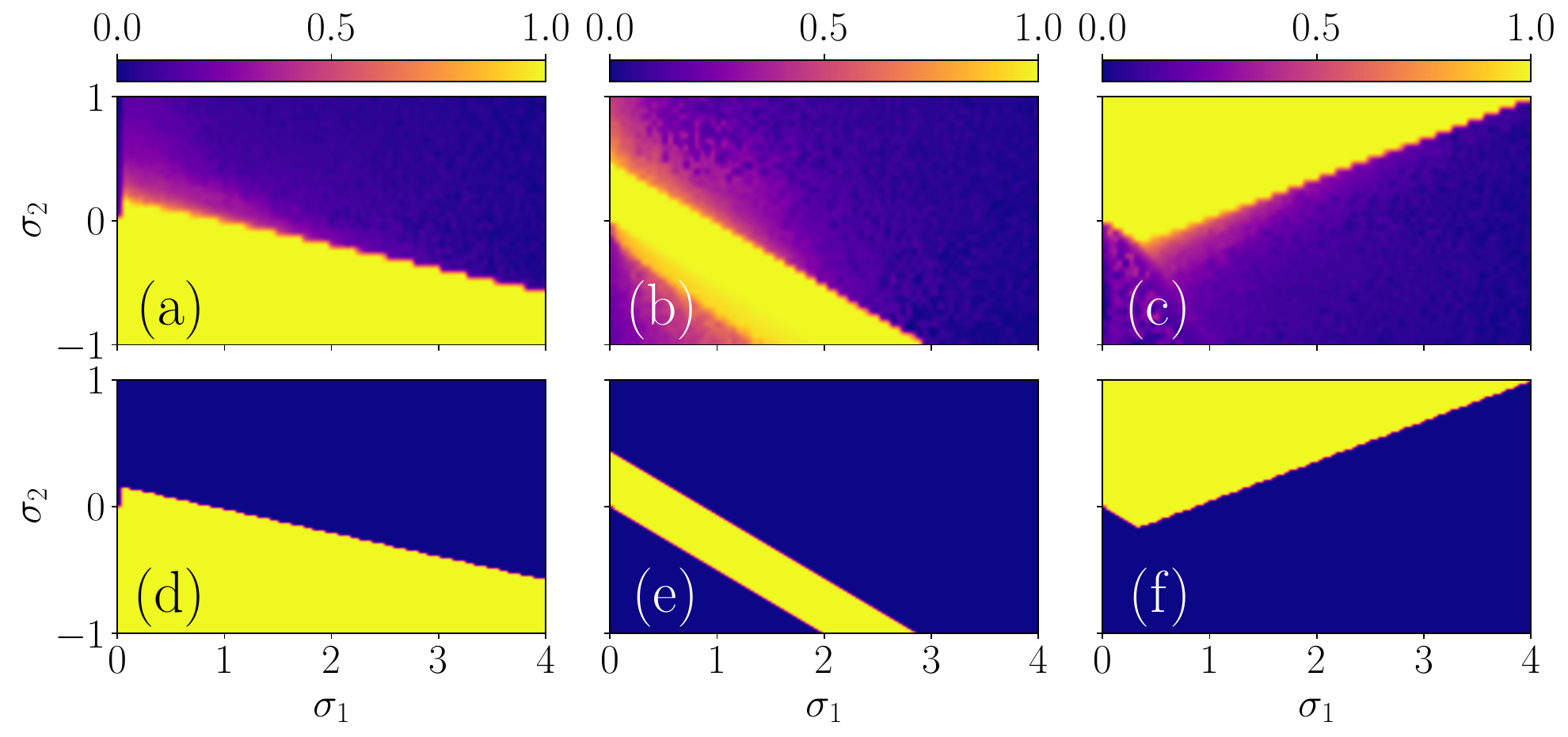}
\caption{
Phase diagrams in the $\sigma_{1}-\sigma_{2}$ plane depict the order parameter starting from a synchronization state with initial conditions close to $\theta_{i}(0) = 0$, $k_{ij}^{(1)}= - a_{ij}^{(1)} \sin \beta_1$ and $k_{ijp}^{(2)}= - a_{ijp}^{(2)} \sin \beta_2$,  for three values of$\beta_2$: (a) $\beta_{2}=0.5\pi$, (b) $\beta_{2}=0.88\pi$, and (c) $\beta_{2}=1.2\pi$ for the case of adaptive higher-order Kuramoto model with lag and an all-to-all coupling. Panels (d)-(f) display the MSF for the adaptive Kuramoto model, wherein the regions with positive MLE are colored blue (dark), while those with negative MLE are yellow (bright). It is observed that the MSF is positive precisely where the order parameter $R_c$ deviates from synchrony, i.e., $R_{c}=1$ The remaining parameters were set to $N=200$, $\alpha_{1}=\alpha_{2}=0.49\pi$, $\beta_{1}=0.88\pi$, $\epsilon_{1}=\epsilon_{2}=0.01$. 
} 
 \label{fig:figall2allhoiadp}
\end{figure*}
The above presented theory will be illustrated by using the adaptive higher-order Kuramoto oscillators with a lag whose equations of motions are given for all $i, j, p\in \{1, 2, \dots, N\}$ by
\begin{align}
\dot \theta_{i} =& \omega - \frac{\sigma_1}{N}\sum_{j=1}^N a_{ij}^{(1)} k_{ij}^{(1)} \sin(\theta_i - \theta_j + \alpha_1) \notag \\ 
& - \frac{\sigma_2}{N^2} \sum_{j=1}^N \sum_{p=1}^N a_{ijp}^{(2)} k_{ijp}^{(2)} \sin(2 \theta_i - \theta_j - \theta_p + \alpha_2), \notag \\
\dot k_{ij}^{(1)} =&-\epsilon_1 [k_{ij}^{(1)} + a_{ij}^{(1)} \sin(\theta_i - \theta_j + \beta_1)], \notag \\
\dot k_{ijp}^{(2)} =& -\epsilon_2 \left[
k_{ijp}^{(2)} + a_{ijp}^{(2)} \sin \left(2 \theta_i - \theta_j - \theta_p + \beta_2 \right) \right]\, . \label{eq:3}
\end{align}
As an example in which the MSE can be further simplified, we consider again the case of globally connected networks. We fixed the number of nodes to $N=200$, the lag parameters $\alpha_1 = 0.49 \pi$, $\beta_1 = 0.88\pi$, $\alpha_2 = 0.49\pi$ and we consider three values $\beta_2 \in\{0.5\pi, 0.88\pi, 1.2\pi\}$. Then, for each value of the latter, we compute the order parameter $R_{c}$ starting from the synchronous state (i.e., $\theta_{i}(0) = 0$, $k_{ij}^{(1)}= - a_{ij}^{(1)} \sin \beta_1$ and $k_{ijp}^{(2)}= - a_{ijp}^{(2)} \sin \beta_2$) in the $(\sigma_{1}, \sigma_{2})$ parameter space where $\sigma_{1}\in [0, 4]$ and $\sigma_{2} \in [-1, 1]$. The results are reported in Fig.~\ref{fig:figall2allhoiadp} where we show (top panels (a)-(c)) the order parameter as a function of $\sigma_1$ and $\sigma_2$ and we compare it with the analytical prediction obtained by solving the MSE (bottom panels (d)-(f)). We can observe an excellent agreement between the two methods indeed in both cases, $R_c\ll 1$ and $\max(\Re\lambda) >0$ (blue areas), testifying the absence of synchronization, do (almost) coincide.

Let us now observe that the higher-order adaptation coupling $\beta_{2}$ in Fig.~\ref{fig:figall2allhoiadp} is chosen in such a way that it satisfies the relations $\beta_{2}< \beta_{1}$, $\beta_{2}= \beta_{1}$, and $\beta_{2} > \beta_{1}$, respectively, while all the other parameter values are kept fixed as in Fig.~\ref{fig:figKlagnohoi} where only the pairwise adaptation is active. We select these values of $\beta_{2}$ to investigate how, depending on the choice of $\beta_{2}$, the stability region is affected by higher-order adaptation, in contrast to constant higher-order interactions.
We observe a relatively opposite scenario in the case of $\beta_{2}<\beta_{1}$ as compared to the result of Fig.~\ref{fig:figKlagnohoi}. The system never settles into the stable synchronization state for larger values of $\sigma_{2}>0$, while for larger negative values of $\sigma_{2}$, the system always remains in the synchronized solution [see Fig.~\ref{fig:figall2allhoiadp} (a), (d)]. For the intermediate values of $\sigma_{2}$, a transition from synchrony to desynchrony emerges with increasing value of $\sigma_{1}$. Fig.~\ref{fig:figall2allhoiadp} (b) and (e) portray the result for $\beta_{2}=\beta_{1}$. Here, we observe that for negative values of $\sigma_{2}$, the system went through a double transition with increasing $\sigma_{1}$, i.e., at first desynchrony to synchrony and then synchrony to desynchrony. Therefore, a significantly large bounded region of stable synchronous solution is observed. While for $\sigma_{2}\ge 0$, a transition from synchrony to desynchrony emerges up to $\sigma_{2} \approx 0.42$, beyond which the system never settles into the stable synchronous solution. Lastly, the results for $\beta_{2}>\beta_{1}$ is illustrated through Fig.~\ref{fig:figall2allhoiadp} (c) and (f), respectively. In this case, the system exhibits qualitatively similar behavior as in the case of no higher-order adaptation, but the region of the stable synchronized state becomes wider.

Therefore, the introduction of higher-order adaptation leads to significantly different qualitative behaviors compared to the case without it. Depending on the choice of higher-order adaptation coupling, synchronization may emerge, be enhanced, weakened, or even completely suppressed.      

\subsection*{Different $\epsilon_1$ and $\epsilon_2$}
\begin{figure}[!ht]
\centering\includegraphics[width=0.87\linewidth]{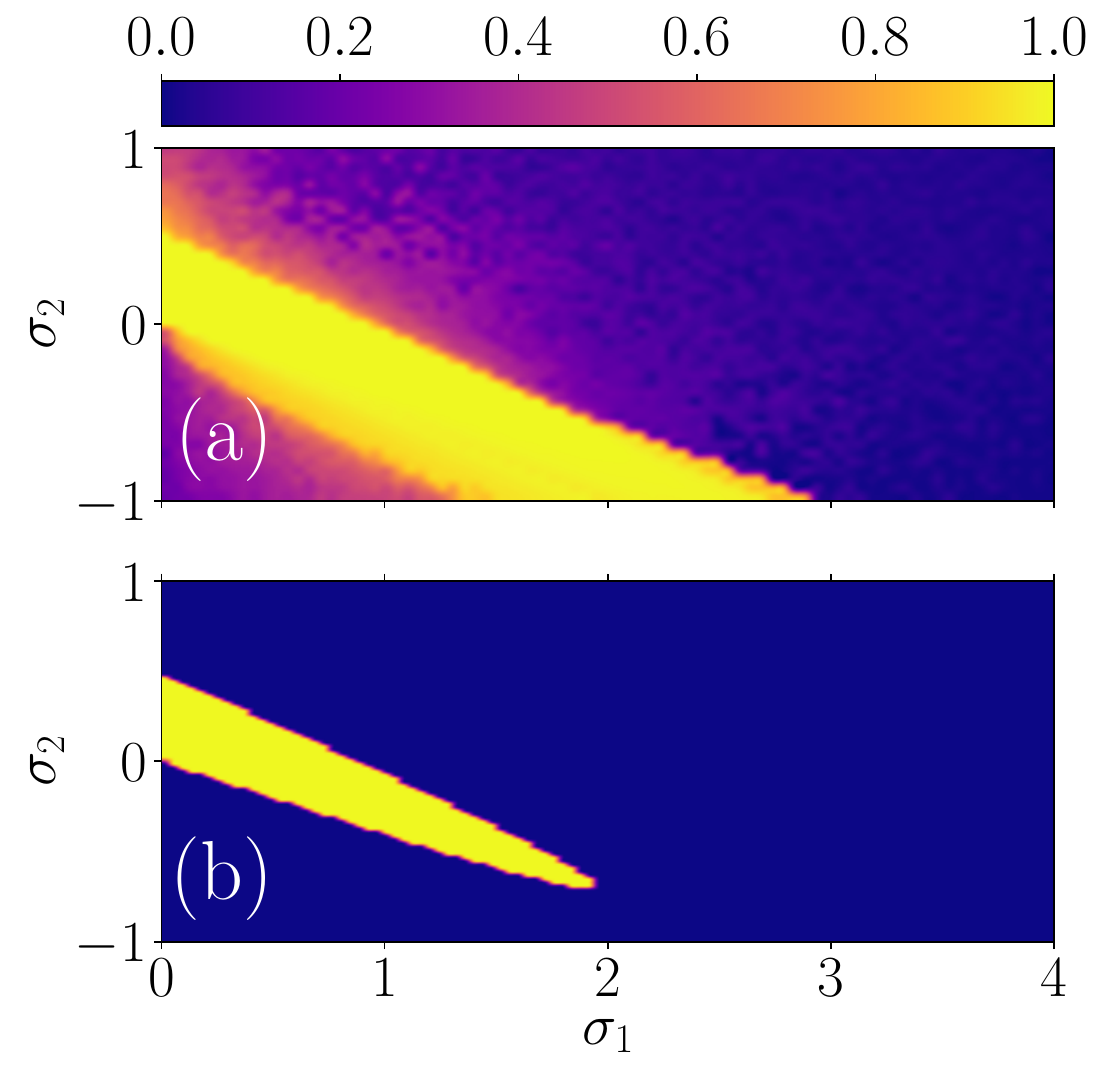}
\caption{
Phase diagrams in the $\sigma_1-\sigma_2$ plane showing (a) the order parameter, numerically computed for $\beta_{2}=0.88\pi$, and (b) the MSF obtained from analytical derivations, presented for the case of an adaptive higher-order Kuramoto model with lag, all-to-all coupling, and $\epsilon_1\neq \epsilon_2$. Regions associated with positive Lyapunov exponents, indicating asynchronous regimes, are colored blue (dark), and the yellow (bright) regions correspond to negative MSF. Other parameters are fixed as in Fig.~\ref{fig:figall2allhoiadp} with $\epsilon_{1}=0.01$ and $\delta=0.001$.
} 
 \label{diff_eps1_eps2}
\end{figure}

As previously mentioned, the preceding analysis with generic coupling functions is only feasible when $\epsilon_{1}=\epsilon_{2}$. However, to what extent can it predict the stable synchronized region when $\epsilon_{1} \neq \epsilon_{2}$? In this regard, here, we demonstrate that the analysis conducted under the assumption $\epsilon_{1}=\epsilon_{2}$ can be applied to provide good results also when the latter condition is not met, namely $\epsilon_{1}-\epsilon_{2}=\delta\neq 0$. In Fig.~\ref{diff_eps1_eps2}, we report the case of $\delta=0.001$, and we can observe that the analytical prediction is capable of determining the synchronization reasonably well and for a large set of parameters $(\sigma_{1}, \, \sigma_{2})$. %

\section{Conclusion}

Summing up, we hereby provide a profound theoretical approach to study collective phenomena, specifically the synchronization, under the combined effect of higher-order interactions and adaptive connectivity. By assuming pairwise and higher-order structures to be regular or by considering noninvasive coupling functions, we have derived the necessary conditions for the stable synchronous solution to exist. Furthermore, we have shown that for two relevant settings, i.e., all-to-all and ring-like, the developed theory resembles the classical MSE approach and thus significantly reduces the intricacy of analytical calculations due to the presence of both adaptivity and higher-order interactions. Finally, our analytical findings have been supported by dedicated numerical simulations, which have fully confirmed the validity of the approach. Additionally, our results highlight how our technique crucially incorporates the fundamental presence of higher-order interactions in adaptive networks, which previously could not be addressed. Our case studies show that the system displays markedly different behaviors when comparing scenarios with and without higher-order adaptation.  

We note that the provided theoretical approach leads to a necessary condition that depends only on the interaction strengths and underlying structural properties of the higher-order structure. Thus, the fact that our approach can be used irrespective of any specific model or choice of coupling functions offers the possibility of extending it to different coupling mechanisms and systems, even those with single or distributed delays \cite{kyrychko2014synchronization}. One of the most realistic examples of a complex system where both higher-order interactions and adaptivity play a pivotal role is the neuronal network. Our generalized approach, therefore, may provide a powerful tool for investigating collective phenomena in neuronal networks, even those involving synaptic plasticity \cite{popovych2015spacing}. Apart from the neuronal networks, adaptation is widely recognized in control theory \cite{delellis2010evolution, yu2012distributed}. Our theoretical approach, therefore, offers a versatile framework for examining various adaptive control schemes across a broad spectrum of dynamical systems, including those with many-body interactions.

\acknowledgments

The work of S.N.J. and P.M. is supported by MoE RUSA 2.0 (Bharathidasan University - Physical Sciences).

\bibliographystyle{apsrev4-2} 
\bibliography{references_new}

\appendix
\section{non-invasive couplings} \label{sec:noninvasive}
Here, we assume the pairwise and three-body coupling functions to be non-invasive, i.e., 
${g}^{(1)}(s, s)=0$ and ${g}^{(2)}(s, s, s)=0$. 

\subsection{Only pairwise adaptation}
In this case, the synchronized solution evolves according to the following equations, 
\begin{subequations}%
\begin{align}
\dot{\mathbf{s}} & = {f}(\mathbf{s}), \\
k_{ij}^{(s)} & = -a_{ij}^{(1)}h(0).
\end{align}%
\end{subequations}%
To analyze the stability of this synchronized state, we slightly perturb the system around the synchronized state with small perturbations $ \mathbf{ {\xi}}_i = \mathbf{ {x}}_i - \mathbf{ {s}} $ and $\chi_{ij} = k_{ij}^{(1)} - k_{ij}^{(s)}$. 
Then, the variational equation can be written as%
\begin{subequations}%
\begin{align}
\dot{\mathbf \xi}_i = & D{f}(\mathbf{s})\xi_i 
- \sigma_1 h(0) \sum_{j=1}^N L_{ij}^{(1)} D_2 {g}^{(1)}(\mathbf{s}, \mathbf{s})\xi_{j} \notag \\ 
& + \sigma_2\sum_{j=1}^N L_{ij}^{(2)}[D_2{g}^{(2)}(\mathbf{s}, \mathbf{s}, \mathbf{s}) 
+ D_3 {g}^{(2)}(\mathbf{s}, \mathbf{s}, \mathbf{s})]\xi_j, \\ 
{ \dot{\chi}}_{ij} = & -\epsilon \left( \chi_{ij} + a_{ij}^{(1)} [Dh(0)(\mathbf{ {\xi}}_i - \mathbf{ {\xi}}_j)]\right).
\end{align}%
\end{subequations}%
In matrix form, the variational equation becomes
\begin{align}
\begin{bmatrix}
\dot{\pmb{\xi}} \\
 \dot{\pmb{\chi}}
\end{bmatrix} 
= 
\begin{bmatrix}
 \mathbf{S} & 0 \\
 -\epsilon \mathbf{C}^{(1)} \otimes Dh(0) & -\epsilon \mathbf{I}_{N^2}
\end{bmatrix}
\begin{bmatrix}
 \pmb{\xi} \\
 \pmb{\chi}
\end{bmatrix} , 
\label{pairnoninvasive}
\end{align}
where
\begin{multline*}
 \mathbf{S} = \mathbf{I}_N \otimes Df(s) - \sigma_1h(0)\mathbf{L}^{(1)} \otimes D_{s} {g}^{(1)} + \sigma_2\mathbf{L}^{(2)} \otimes D_{s} {g}^{(2)}, 
 \end{multline*}
with $D_{s} {g}^{(1)}=D_2g^{(1)}(s, s)$, and $D_{s} {g}^{(2)}= D_2g^{(2)}(s, s, s) + D_3g^{(2)}(s, s, s)$, respectively. $\mathbf{C}^{(1)}$ is a constant matrix of order $N^2 \times N$ (details are provided in the Appendix \ref{sec:msf_calculation}).

To determine the stability condition for the associated synchronous solution, one needs to solve the system of equations \eqref{pairnoninvasive} to calculate the maximum Lyapunov exponents. The form of Eq.~\eqref{pairnoninvasive} implies that the Jacobian matrix possesses $N^2$ numbers of $-\epsilon$ eigenvalues. Furthermore, it is a lower diagonal block matrix. Hence, the stability of $\mathbf{S}$, i.e., finding the maximum Lyapunov exponents by solving $\dot {\pmb{\xi}}=\mathbf{S} \pmb{\xi}$ provides the necessary condition for the stable synchronized solution, subject to $\epsilon>0$. Now, $\dot {\pmb{\xi}}=\mathbf{S} \pmb{\xi}$ is a coupled linear differential equation of dimension $Nd$. To further simplify it, we project the projection variables $\xi$ onto the eigenspace of the pairwise Laplacian matrix $\mathbf{L}^{(1)}$ by introducing a new set of variables $\pmb{\zeta}=(\mathbf{U} \otimes \mathbf{I}_{d})^{\top} \pmb{\xi}$. Here, $\mathbf{U}$ is a $N\times N$ matrix whose columns are the orthonormal eigenvectors of the Laplacian $\mathbf{L}^{(1)}$, i.e., $\mathbf{U}^{\top}\mathbf{L}^{(1)}\mathbf{U}=\mathrm{diag} \{0= \mu^{(1)}_{1}, \mu^{(1)}_{2}, \dots, \mu^{(1)}_{N} \}$. Using this new coordinate system, the variational equation becomes 
\begin{align}
 \dot{\zeta}_{i} = [Df(s) - \sigma_{1}h(0)\mu_{i}^{(1)}D_sg^{(1)}] \zeta_{i} 
 + \sigma_{2}\sum\limits_{j=1}^{N} \tilde{L}_{ij}^{2}D_sg^{(2)}\zeta_{j}, 
\end{align}
where we introduced $\tilde{{L}}^{(2)}={U}^{\top}{L}^{(2)}{U}$ whose entries on the first row and column are all zero. Thus, $\zeta_{1}$ corresponds to the modes parallel to the synchronous solution, and $\zeta_{i}, \; i=2, 3, \dots, N$ is associated with the transverse modes. In general, this transverse variation equation can not decouple further. However, when the higher-order Laplacian is a scalar multiple of the pairwise one and in the scenario of commutating Laplacian matrices, the variational equation can be fully decoupled into $N$ numbers of $d$-dimensional equation. 
\subsection{Both pairwise and higher-order adaptation}
In this case, the synchronized solution follows
\begin{subequations}
\begin{align}
\dot{\bf{s}} & = {f}(\mathbf{s}), \\ 
k_{ij}^{(1),s} & = -a_{ij}^{(1)}{h}^{(1)}(0), \\
k_{ijp}^{(2),s} & = -a_{ijp}^{(2)}{h}^{(2)}(0). 
\end{align}
\end{subequations}
The corresponding variation equation can be written as, 
\begin{subequations}
 \label{}
\begin{align}
\dot{\mathbf{\xi}}_i & = D{f}(\mathbf{s})\mathbf{\xi}_i
 -\sigma_1 {h}^{(1)}(0)\sum_{j=1}^N {L}_{ij}^{(1)} {D}_2{g}^{(1)}(\mathbf{s}, \mathbf{s})\xi_j \notag \\ 
 - &\sigma_2{h}^{(2)}(0)\sum_{j=1}^N {L}_{ij}^{(2)}({D}_2{g}^{(2)}(\mathbf{s}, \mathbf{s}, \mathbf{s})+
{D}_3{g}^{(2)}(\mathbf{s}, \mathbf{s}, \mathbf{s}))\xi_j \notag \\
 - &\sigma_1 \sum_{j=1}^N a_{ij}^{(1)} \chi_{ij} {g}^{(1)}(\mathbf{s}, \mathbf{s}) 
- \sigma_2\sum_{j=1}^N \sum_{p=1}^N a_{ijp}^{(2)}\eta_{ijp}{g}^{(2)}(\mathbf{s}, \mathbf{s}, \mathbf{s}), 
\end{align}
\begin{align}
 \dot{\chi}_{ij} &= - \epsilon_1(\chi_{ij} + a_{ij}^{(1)}{Dh}^{(1)}(0)(\xi_i - \xi_j)), \\
 \dot{\eta}_{ijp} &= - \epsilon_2(\eta_{ijp} + a_{ijp}^{(2)}{Dh}^{(2)}(0)(2\xi_i - \xi_j - \xi_p)).
\end{align}
\end{subequations}
Then, in matrix form, the variational equations can be written as
\begin{align}
\scriptsize
\begin{bmatrix}
 \dot{\pmb{\xi}} \\
 \dot{\pmb{\chi}} \\
 \dot{\eta}
\end{bmatrix} 
= 
\begin{bmatrix}
 \mathbf{S}_2 & 0 & 0 \\
 -\epsilon_{1} \mathbf{C}^{(1)} \otimes Dh^{(1)}(0) & -\epsilon_{1} \mathbf{I}_{N^2} & 0 \\
 -\epsilon_{2} \mathbf{C}^{(2)} \otimes Dh^{(2)}(0) & 0 & -\epsilon_{2} \mathbf{I}_{N^3} 
\end{bmatrix}
\begin{bmatrix}
 \pmb{\xi}\\
 \pmb{\chi} \\
 \pmb{\eta}
\end{bmatrix} 
\end{align}
where
\begin{multline*}
 \mathbf{S}_ 2 = \mathbf{I}_N \otimes Df(s) 
 - \sigma_1h^{(1)}(0)\mathbf{L}^{(1)} \otimes D_{s} {g}^{(1)} \\ - \sigma_2h^{(2)}(0)\mathbf{L}^{(2)} \otimes D_{s} {g}^{(2)}. 
\end{multline*}
To determine the stability condition for the associated synchronous solution, one needs to solve the above system of variation equations to calculate the maximum Lyapunov exponents. The form of the equation implies that the Jacobian matrix possesses $N^2$ numbers of $-\epsilon_{1}$ eigenvalues and $N^3$ numbers of $\epsilon_{2}$ eigenvalues. Furthermore, it is a lower diagonal block matrix. Hence, the stability of $\mathbf{S}_2$, i.e., finding the maximum Lyapunov exponents by solving 
$\dot{\pmb{\xi}}=\mathbf{S}_2 \; \pmb{\xi}$ provides the necessary condition for the stable synchronized solution, subject to $\epsilon_1, \epsilon_2>0$. Further simplification of the variation equation $\dot{\pmb{\xi}}=\mathbf{S}_2 \; \pmb{\xi}$ can be done similarly to the previous case. Hence, in this case, we can analytically predict the stability of the synchronous solution even if $\epsilon_{1}$ and $\epsilon_{2}$ are different. 
\begin{figure}[!ht]
\centering\includegraphics[width=0.9\linewidth]{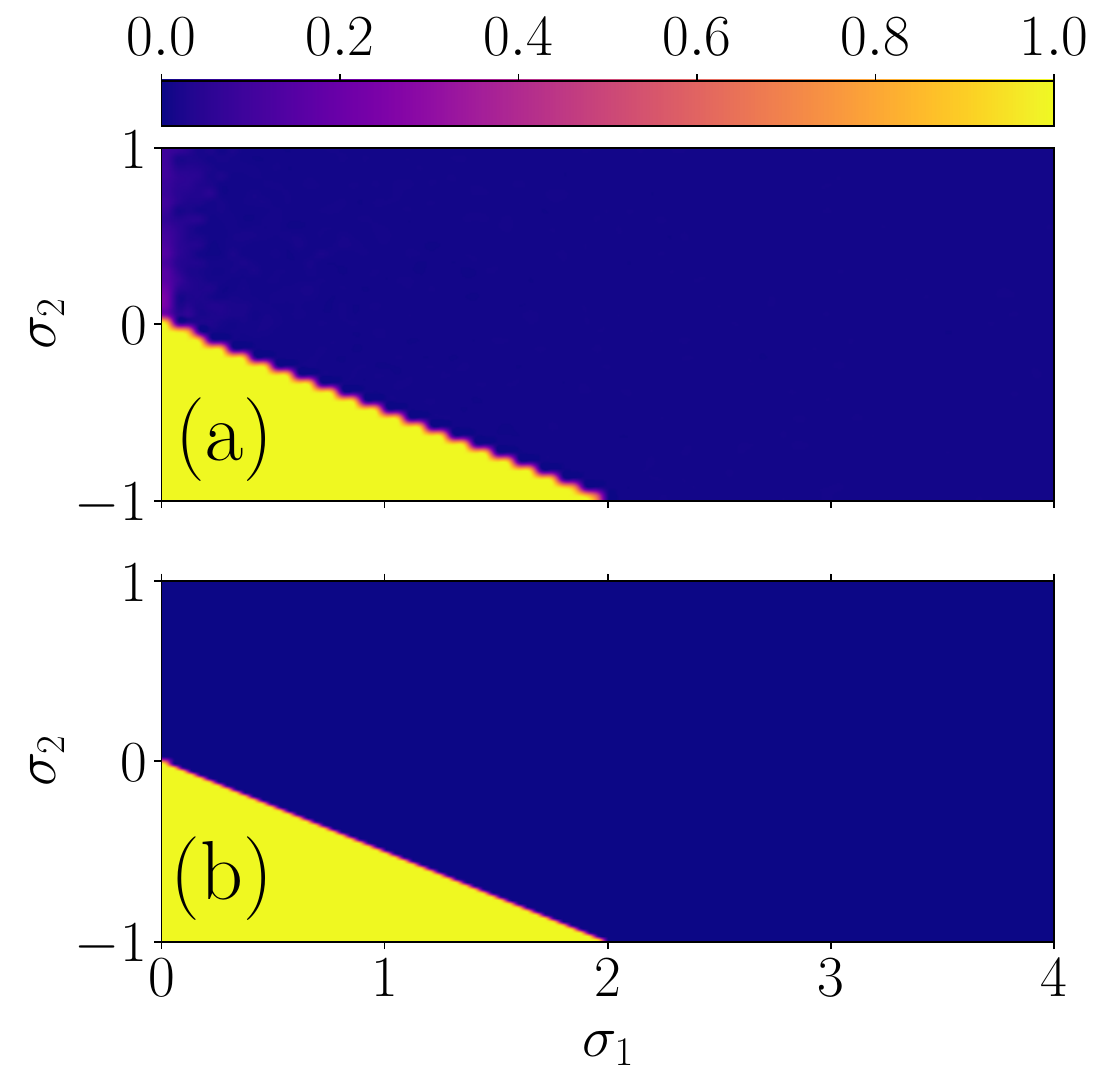}
\caption{
Phase diagrams in the $\sigma_1-\sigma_2$ plane showing  (a) the order parameter, numerically computed for $\beta_{2}=0.88\pi$, and (b) the MSF obtained from analytical derivations, for the case of a higher-order Kuramoto model with lag, all-to-all topology having both pairwise and higher-order adaptation, and noninvasive coupling schemes. Regions associated with positive Lyapunov exponents, indicating asynchronous regimes, are colored blue (dark), whereas yellow (bright) represents the region of negative Lyapunov exponents. Other parameters are fixed as follows: $N=100$, $\alpha_{1}=\alpha_{2}= 0$, $\beta_{1}=\beta_{2}= 0.88\pi$, $\epsilon_{1}=0.01$, $\epsilon_{2}=0.02$.}
\label{noninvasive}
\end{figure}

To show that the presented theory corresponding to the noninvasive coupling works even for $\epsilon_{1} \neq \epsilon_{2}$, we consider an ensemble of $N=100$ globally coupled phase-lagged Kuramoto oscillators given by Eq.~\eqref{eq:3}. Since the coupling functions $g^{(i)}$, $i=1, 2$ are synchronization non-invasive, we consider the lag parameters $\alpha_{i}=0$. The time scale separation parameters are taken as $\epsilon_{1}=0.01$ and $\epsilon_{2}=0.02$. All the other parameters are kept fixed at a nominal value. The results are reported in Fig.~\ref{noninvasive} where the MSF \textcolor{blue}{[cf. Fig.~\ref{noninvasive}(b)]} is computed as a function of $\sigma_1$ and $\sigma_2$ and compared with the order parameter $R_c$ \textcolor{blue}{[cf. Fig.~\ref{noninvasive}(a)]}. It is observable that the analytical prediction can determine the region of stable synchronous solution very well. 

\section{Derivation of the master stability equations} \label{sec:msf_calculation}

Here, we provide a detailed, step-by-step derivation of the master stability equations discussed in the main text. 

\subsection{Derivation of the master stability equation for the system with pairwise adaptation and constant higher-order interactions} \label{sec:nohoiadaptsupply}

The general dynamical equations of the adaptive higher-order networks with adaptation only in pairwise interactions are given by%
\begin{subequations}%
\begin{align}
\mathbf{ \dot{x}}_i = & {f}(\mathbf{x}_i) - \sigma_1 \sum_{j=1}^N a_{ij}^{(1)} k_{ij}^{(1)}(t) {g}^{(1)}(\mathbf{x}_i,\mathbf{x}_j) \notag  \\ & -\sigma_2{\sum_{j=1}^N}{\sum_{p=1}^N} a_{ijp}^{(2)} k_{ijp}^{(2)}{g}^{(2)}(\mathbf{x}_i,\mathbf{x}_j,\mathbf{x}_p),  \\
\dot k_{ij}^{(1)} & = -\epsilon[k_{ij}^{(1)} + a_{ij}^{(1)}{h}(\mathbf{x}_i-\mathbf{x}_j)],
\end{align}%
\end{subequations}%
where the adjacency matrix (tensor) satisfies the constant degree property for each node, i.e., $\sum a_{ij}^{(1)} = r^{(1)}$ and $\sum_{j=1}^N{\sum_{p=1}^N} a_{ijp}^{(2)} = 2r^{(2)}$.
Let us observe that $(\mathbf{s},k^{(s)})$ is the synchronized solution for the adaptive dynamical system when it satisfies the following evolution equations: %
\begin{subequations}%
\begin{align}
& \mathbf{ \dot{s}} =   {f}(\mathbf{s}) + \sigma_1 r^{(1)}h(0) {g}^{(1)}(\mathbf{s},\mathbf{s}) -2\sigma_2r^{(2)}{g}^{(2)}(\mathbf{s},\mathbf{s},\mathbf{s}), \\
& k_{ij}^{(s)}  =  -a_{ij}^{(1)}h(0).
\end{align}%
\end{subequations}%
Now, to analyze the stability of this synchronized state, we slightly perturb the system around the synchronized state with the perturbations defined as $ \mathbf{ {\xi}}_i = \mathbf{ {x}}_i - \mathbf{ {s}} $ and $\chi_{ij} =  k_{ij}^{(1)} -  k_{ij}^{(s)}$. Using Taylor series expansion to linearize the system around the synchronized state, we can obtain the variational equations as follows, %
\begin{subequations}%
\begin{widetext}
\begin{align}
\dot{\mathbf \xi}_i = & D{f}(\mathbf{s})\xi_i  - \sigma_1 \sum_{j=1}^N a_{ij}^{(1)}{g}^{(1)}(\mathbf{s},\mathbf{s}) \chi_{ij} + \sigma_1r^{(1)}h(0)D_1 {g}^{(1)}(\mathbf{s},\mathbf{s})\xi_i + \sigma_1h(0)\sum_{j=1}^Na_{ij}^{(1)} D_2 {g}^{(1)}(\mathbf{s},\mathbf{s})\xi_j \notag \\ &   - 2\sigma_2r^{(2)} D_1 {g}^{(2)}(\mathbf{s},\mathbf{s},\mathbf{s})\xi_i -\sigma_2\sum_{j=1}^N{\sum_{p=1}^N} a_{ijp}^{(2)}[D_2 {g}^{(2)}(\mathbf{s},\mathbf{s},\mathbf{s})\xi_j + D_3 {g}^{(2)}(\mathbf{s},\mathbf{s},\mathbf{s})\xi_p], 
\end{align}
\end{widetext}
\begin{align}
{ \dot{\chi}}_{ij} & = -\epsilon ( \chi_{ij} + a_{ij}^{(1)} [Dh(0)({ {\xi}_j} - { {\xi}_i})],
\end{align}%
\end{subequations}%
where $Df$ and $Dh^{(1)}$ are the Jacobians of $f$ and $h$, $D_{i}g^{(j)}$ denotes the Jacobian of the function $g^{j}$, $(j=1,2)$, with respect to the $i$-th variables, $i=1,2,3$. 

By introducing the pairwise and higher-order Laplacians $\mathbf{L}^{(1)}$ and $\mathbf{L}^{(2)}$, defined as
\begin{align}
    \begin{array}{l}
         L^{(1)}_{ij}=\begin{cases}
             -a^{(1)}_{ij}, & i \neq j \\
             \sum\limits_{j=1}^{N} a^{(1)}_{ij}=r^{(1)}, & i=j 
         \end{cases} 
    \end{array}
\end{align}
and, 
\begin{align}
    \begin{array}{l}
         L^{(2)}_{ij}=\begin{cases}
             -\sum\limits_{k=1}^{N} a^{(2)}_{ijk}, & i \neq j \\
             \sum\limits_{j=1}^{N}\sum\limits_{k=1}^{N} a^{(2)}_{ijk}=2r^{(2)}, & i=j 
         \end{cases} 
    \end{array}
\end{align}
we can rewrite the variational equation as, 
\begin{widetext}
\begin{align}
\dot{\mathbf \xi}_i = & D{f}(\mathbf{s})\xi_i  - \sigma_1 \sum_{j=1}^N a_{ij}^{(1)} {g}^{(1)}(\mathbf{s},\mathbf{s}) \chi_{ij}   
+ \sigma_1r^{(1)}h(0)[D_1 {g}^{(1)}(\mathbf{s},\mathbf{s}) + D_2 {g}^{(1)}(\mathbf{s},\mathbf{s})]\xi_i \notag \\ &   
- \sigma_1 h(0) \sum_{j=1}^N L_{ij}^{(1)} D_2 {g}^{(1)}(\mathbf{s}, \mathbf{s})\xi_{j} - 2\sigma_2r^{(2)} [D_1 {g}^{(2)}(\mathbf{s},\mathbf{s},\mathbf{s}) + D_2 {g}^{(2)}(\mathbf{s},\mathbf{s},\mathbf{s}) + D_3 {g}^{(2)}(\mathbf{s},\mathbf{s},\mathbf{s})] \xi_i  \notag \\ &   
+ \sigma_2\sum_{j=1}^N L_{ij}^{(2)}[D_1 {g}^{(2)}(\mathbf{s},\mathbf{s},\mathbf{s}) + D_2 {g}^{(2)}(\mathbf{s},\mathbf{s},\mathbf{s})]\xi_j 
\end{align}
\end{widetext}
and 
\begin{align}
{ \dot{\chi}}_{ij} & = -\epsilon ( \kappa_{ij} + a_{ij} [Dh(0)({ {\xi}_j} - { {\xi}_i})].
\end{align}

Now, to write the variational equation in matrix form, we introduce the vectorized form for the perturbations as ${\xi} = \mathbf{x}- I_N \otimes \mathbf{s} $ and $ {\chi} = k - k^s $ with
\begin{align}
    \mathbf{x} = & (\mathbf{x}_1^{\top}, \dots ,\mathbf{x}_N^{\top} )^{\top},  \notag \\   
    k & =  (k_{11},\dots,k_{1N},k_{21},\dots,k_{2N}, \dots, k_{N1},\dots, k_{NN})^{\top},
\end{align}
where we stack the rows of $k_{ij}$ successively to make the $N\times N$ matrix $\chi_{ij}$ to a $N^2$ dimensional vector. 

Thereafter, to proceed further we introduce few notations and matrices as follows. We consider
\begin{align}
    \mathbf{a}_i^{(1)} = (a_{i1}^{(1)},\dots,a_{iN}^{(1)}),  \;\;\;
\mathrm{diag}(\mathbf{a}_i^{(1)}) = 
\begin{pmatrix}
    a_{i1}^{(1)}& & \\
     &\dots& \\
     & & a_{iN}^{(1)}  \notag \\   
\end{pmatrix}
\end{align}
and the $N \times N^2$, $N^2 \times N$ matrices as,
\begin{align}
\mathbf{B}^{(1)} = 
\begin{pmatrix}
    \mathbf{a}_{1}^{(1)}& & \\
     &\dots& \\
     & & \mathbf{a}_{N}^{(1)}  \notag \\   
\end{pmatrix}, \;\;\;
\mathbf{D}^{(1)} = 
\begin{pmatrix}
    \mathrm{diag}(\mathbf{a}_1^{(1)}) \\
     \vdots \\
     \mathrm{diag}(\mathbf{a}_N^{(1)})  \notag \\   
\end{pmatrix}.
\end{align}
We further can construct the $\mathbf{C}^{(1)}$ matrix of dimension $ N^2 \times N$ from $\mathbf{B}^{(1)}$ and $\mathbf{D}^{(1)}$ as, 
\begin{align}
    \mathbf{C}^{(1)} = B^{(1)^{\top}} - \mathbf{D}^{(1)}
\end{align}
Using all the above notations and matrices, the variational equation can be written in block matrix form as,
\begin{align}
\begin{pmatrix}
        \dot{\pmb{\xi}} \\
        \dot{\pmb{\chi}}
\end{pmatrix} 
= 
\begin{pmatrix}
        \mathbf{S} & -\sigma_1 \mathbf{B}^{(1)} \otimes {g}^{(1)}(\mathbf{s},\mathbf{s}) \\
        -\epsilon \mathbf{C}^{(1)} \otimes Dh(0) & -\epsilon \mathbf{I}_{N^2}
\end{pmatrix}
\begin{pmatrix}
        \pmb{\xi} \\
        \pmb{\chi}
\end{pmatrix} ,
\end{align}
where $\mathbf{S} = \mathbf{I}_N \otimes Df(s) + \sigma_1h(0)[r\mathbf{I}_N \otimes ( D_1g^{(1)} + D_2g^{(1)}) ] - 2\sigma_2[ r^{(2)}\mathbf{I}_N  \otimes [D_1g^{(2)} + D_2g^{(2)} +  D_3g^{(2)}] - \sigma_1h(0)\mathbf{L}^{(1)} \otimes D_2g^{(1)} + \sigma_2 \mathbf{L}^{(2)} \otimes [ D_2g^{(2)} + D_3g^{(2)}]$. $\mathbf{I}_N$ is the $N\times N$ identity matrix, $\mathbf{B}^{(1)}$ and $\mathbf{C}^{(1)}$ satisfy the relation $\mathbf{B}^{(1)}\mathbf{B}^{(1)^{\top}}=r^{(1)}\mathbf{I}_N$ and $\mathbf{B}^{(1)}\mathbf{C}^{(1)}=\mathbf{L}^{(1)}$. 

From the structure of the variational equation, one can easily find that it has $(N^2-N)$ eigenvalues $\lambda = -\epsilon$. The eigenspace corresponding to those eigenvalues are time-independent and can be obtained from, 
\begin{align}
\begin{pmatrix}
        \mathbf{S} + \epsilon \mathbf{I}_{Nd} & -\sigma_1  \mathbf{B}^{(1)} \otimes {g}^{(1)}(\mathbf{s},\mathbf{s}) \\
         -\epsilon \mathbf{C}^{(1)} \otimes Dh(0) & 0
\end{pmatrix}
\begin{pmatrix}
        \pmb{\xi} \\
        \pmb{\chi}
\end{pmatrix} 
=
0.
\end{align}
From this it is obvious that $(\pmb{\xi},\pmb{\chi})$ satisfying $\pmb{\xi}=0$ and $\mathbf{B}^{(1)}\pmb{\chi}=0$ are the time-independent eigenvectors. Now, since $\pmb{\chi}$ is $N^2$ dimensional and $\text{rank}(\mathbf{B}^{(1)}) = N$ when the row sum $r^{(1)}$ is nonzero, we can say that there exists $(N^2-N)$ linearly independent eigenvectors which spans the eigenspace corresponding to the eigenvalue $-\epsilon$.

For the $(N^2 - N)$ eigenvalues, there exists  $(N^2 - N)$ independent eigenvectors $v_l (l = 1, 2, \ldots ,N^2-N)$ spanning the kernal of $\mathbf{B}^{(1)}$. By using the Gram-Schmidt procedure, we can find the orthonormal basis for the kernel of $\mathbf{B}^{(1)}$ as $\mathrm{ker}(\mathbf{B}^{(1)}) = \mathrm{span}(y_1, \ldots,y_{N^2-N})$.
With this basis, we can define matrices $\mathbf{R}^{(1)}$ and $\mathbf{Q}^{(1)}$ of dimensions $N^2 \times (N^2 - N)$ and $(N^2 +Nd) \times (N^2 + Nd)$, respectively as 
$\mathbf{R}^{(1)} = (y_1, y_2, \ldots ,y_{N^2-N}) $ and $\mathbf{Q}^{(1)}$,
\begin{align}
\mathbf{Q}^{(1)}
=
\begin{pmatrix}
           \mathbf{I}_{Nd} & 0 & 0 \\
           0& (1/r^{(1)})\mathbf{B}^{(1)^{\top}} & \mathbf{R}^{(1)}
\end{pmatrix}.
\end{align}
The left inverse of $\mathbf{Q}^{(1)}$ is
\begin{align}
\mathbf{Q}^{(1)^{-1}}
=
\begin{pmatrix}
           \mathbf{I}_{Nd} & 0 \\
           0& \mathbf{B}^{(1)} \\
           0& \mathbf{R}^{(1)}
\end{pmatrix},
\end{align}
So that, $\mathbf{Q}^{(1)^{-1}}\mathbf{Q}^{(1)} = \mathbf{I}_{N^2+Nd}$. 
The variational equation can be written as,
\begin{widetext}
\begin{align}
\begin{pmatrix}
        \dot{\pmb{\xi}} \\
        \dot{\pmb{\chi}}
\end{pmatrix} 
=
\mathbf{Q}^{(1)^{-1}}
\begin{pmatrix}
        \mathbf{S} & -\sigma_1 \mathbf{B}^{(1)} \otimes {g}^{(1)}(\mathbf{s},\mathbf{s}) \\
        -\epsilon \mathbf{C}^{(1)} \otimes Dh(0) & -\epsilon \mathbf{I}_{N^2}
\end{pmatrix}
\mathbf{Q}^{(1)}
\begin{pmatrix}
        \pmb{\xi} \\
        \pmb{\chi}
\end{pmatrix}, 
\end{align}
where with a slight abuse we use the same letter to indicate the new transformed co-ordinates. Now, 
\begin{align}
\mathbf{Q}^{(1)^{-1}}
\begin{pmatrix}
        \mathbf{S} & -\sigma_1 \mathbf{B}^{(1)} \otimes {g}^{(1)}(\mathbf{s},\mathbf{s}) \\
        -\epsilon \mathbf{C}^{(1)} \otimes Dh(0) & -\epsilon \mathbf{I}_{N^2}
\end{pmatrix}
\mathbf{Q}^{(1)}
=  &\,
\mathbf{Q}^{(1)^{-1}}
\begin{pmatrix}
        \mathbf{S} & -\sigma_1 \mathbf{B}^{(1)} \otimes {g}^{(1)}(\mathbf{s},\mathbf{s}) \\
        -\epsilon \mathbf{C}^{(1)} \otimes Dh(0) & -\epsilon \mathbf{I}_{N^2}
\end{pmatrix} \notag \\
& \, \times
\begin{pmatrix}
           \mathbf{I}_{Nd} & 0 & 0 \\
           0& (1/r^{(1)})\mathbf{B}^{(1)^{\top}} & \mathbf{R}^{(1)}
\end{pmatrix} \notag \\
= & \,
\mathbf{Q}^{(1)^{-1}}
\begin{pmatrix}
        \mathbf{S} & -\sigma_1 \mathbf{I}_N \otimes {g}^{(1)}(\mathbf{s},\mathbf{s}) & 0 \\
        -\epsilon \mathbf{C}^{(1)} \otimes Dh(0) & -\epsilon /r^{(1)} \mathbf{B}^{(1)^{\top}} & -\epsilon \mathbf{R}^{(1)}
\end{pmatrix} \notag \\
= & \, \begin{pmatrix}
           \mathbf{I}_{Nd} & 0 \\
           0& \mathbf{B}^{(1)} \\
           0& \mathbf{R}^{(1)}
\end{pmatrix}
\begin{pmatrix}
        \mathbf{S} & -\sigma_1 \mathbf{I}_N \otimes {g}^{(1)}(\mathbf{s},\mathbf{s}) & 0 \\
        -\epsilon \mathbf{C}^{(1)} \otimes Dh(0) & -\epsilon /r^{(1)} \mathbf{B}^{(1)^{\top}} & -\epsilon \mathbf{R}^{(1)}
\end{pmatrix} \notag \\
= & \,
\begin{pmatrix}
        \mathbf{S} & -\sigma_1 \mathbf{I}_N \otimes {g}^{(1)}(\mathbf{s},\mathbf{s}) & 0 \\
        -\epsilon \mathbf{L}^{(1)} \otimes Dh(0) & -\epsilon \mathbf{I}_N & 0 \\
        -\epsilon \mathbf{R}^{(1)^{\top}}\mathbf{C}^{(1)} \otimes Dh(0) & 0 & -\epsilon \mathbf{I}_{N^2-N}
\end{pmatrix} \notag
\end{align}
\end{widetext}
From this, we can obtain $(N + Nd)$ coupled master equations as
\begin{align}
\begin{pmatrix}
        \dot{\pmb{\xi}} \\
        \dot{\pmb{\chi}_M}
\end{pmatrix} 
=
\begin{pmatrix}
 \mathbf{S} & -\sigma_1 \mathbf{I}_N \otimes {g}^{(1)}(\mathbf{s},\mathbf{s}) \\
        -\epsilon \mathbf{L}^{(1)} \otimes Dh(0) & -\epsilon \mathbf{I}_N \\
\end{pmatrix}
\begin{pmatrix}
        \pmb{\xi} \\
         \pmb{\chi}_M
\end{pmatrix},
\end{align}
where $\pmb{\chi}_M = \pmb{\chi}_1$. Moreover, we yield $N^2 - N$ salve equations which can be solved explicitly once the variables associated with the master equations are known. This equations are given by
\begin{align}
    \dot{\pmb{\chi}}_S 
    =
\begin{pmatrix}
    -\epsilon \mathbf{R}^{(1)^{\top}}\mathbf{C}^{(1)} \otimes Dh(0) & 0 & -\epsilon \mathbf{I}_{N^2-N}
\end{pmatrix}
\begin{pmatrix}
        {\pmb{\xi}} \\
        {\pmb{\chi}}_M \\
         {\pmb{\chi}_S}
\end{pmatrix},
\end{align}
where ${\pmb{\chi}}_S = ({{\chi}}_2^{\top}, {{\chi}}_3^{\top}, \ldots ,{{\chi}}_N^{\top})^{\top}$. Now, corresponding to the zero row-sum symmetric Laplacian matrices there exists the matrices with its columns as orthonormal eigenvectors that diagonalize the Laplacian matrices. For example, there exists the matrix $\mathbf{U}$ such that $\mathbf{U}^{\top}\mathbf{L}^{(1)}\mathbf{U}=\mathbf{D}$, where $\mathbf{D}$ is the diagonal matrix with the diagonal entries being the eigenvalues of $\mathbf{L}^{(1)}$. In order to use this relation to decouple the master variational equation, we introduce a change of co-ordinates defined as
\begin{align}
\begin{pmatrix}
    \mathbf{U} \otimes \mathbf{I}_d & 0 \\
    0 & \mathbf{U} 
\end{pmatrix}
\begin{pmatrix}
    \pmb{\xi}\\
    \pmb{\chi}_M     
\end{pmatrix}
=
\begin{pmatrix}
    \pmb{\zeta} \\
    \pmb{\eta}
\end{pmatrix}.
\end{align}
This yields
\begin{align}
\begin{pmatrix}
    \dot{\pmb{\zeta}} \\
   \dot{\pmb{\eta}}
\end{pmatrix} =
\begin{pmatrix}
        \tilde{\mathbf{S}} & -\sigma_1 \mathbf{I}_N \otimes {g}^{(1)}(\mathbf{s},\mathbf{s}) \\
        -\epsilon \mathbf{D}_L^{(1)} \otimes Dh(0) & -\epsilon \mathbf{I}_{N}
\end{pmatrix}
\begin{pmatrix}
    \pmb{\zeta} \\
    \pmb{\eta}
\end{pmatrix},
\end{align}
where $ \tilde{\mathbf{S}} = \mathbf{U}^{\top}\mathbf{S}\mathbf{U}= \mathbf{I}_N \otimes Df(s) + \sigma_1h(0)[r^{(1)} \mathbf{I}_N \otimes ( D_1g^{(1)} + D_2g^{(1)}) ] - 2\sigma_2[ r^{(2)}\mathbf{I}_N  \otimes [D_1g^{(2)} + D_2g^{(2)} +  D_3g^{(2)}] - \sigma_1h(0)\mathbf{D} \otimes D_2g^{(1)} + \sigma_2\mathbf{U}^H \mathbf{L}^{(2)} \mathbf{U}\otimes [ D_2g^{(2)} + D_3g^{(2)}]$. In explicit form it can be expressed as,%
\begin{subequations}%
\begin{align}
\dot \zeta_{i}= & [Df(s) + \sigma_1h(0)r^{(1)} D {g}^{(1)} - \sigma_1h(0) \mu^{(1)}_{i} D_{s} {g}^{(1)} \notag \\ & - 2\sigma_2 r^{(2)} D {g}^{(2)}]\zeta_{i} +\sum\limits_{j=1}^{N} \Tilde{L}^{2}_{ij}D_{s} \mathbf{g}^{(2)}\zeta_{j}-\sigma_1g^{(1)}(s,s)\eta_{i},
\\
\dot \eta_{i} = & -\epsilon[ \mu^{(1)}_{i} Dh(0) \zeta_{i} + \eta_{i} ],
\label{genmse1B}
\end{align}%
\end{subequations}%
with $\tilde{\mathbf{L}}^{(2)}=\mathbf{U}^{\top}\mathbf{L}^{(2)}\mathbf{U}$. This is the required master stability equation.

\subsection{Derivation of the master stability equation for the system with both pairwise and higher-order adaptations  }

The general dynamical equations of the adaptive weighted simplicial complexes with adaptation in both pairwise and higher-order interactions are %
\begin{subequations}%
\begin{align}
\dot{\mathbf{x}}_i =& {f}(\mathbf{x}_i) - \sigma_1 \sum_{j=1}^N a_{ij}^{(1)} k_{ij}^{(1)}(t) {g}^{(1)}(\mathbf{x}_i,\mathbf{x}_j) \notag \\ & -\sigma_2{\sum_{j=1}^N}{\sum_{p=1}^N} a_{ijp}^{(2)} k_{ijp}^{(2)}(t) {g}^{(2)}(\mathbf{x}_i,\mathbf{x}_j,\mathbf{x}_p)\, , \\
\label{eq:2:2}
\dot k_{ij}^{(1)} = & -\epsilon_1[k_{ij}^{(1)} + a_{ij}^{(1)} {h}^{(1)}(\mathbf{x}_i-\mathbf{x}_j)], \\
\dot k_{ijp}^{(2)} = &-\epsilon_2[k_{ijp}^{(2)} + a_{ijp}^{(2)} {h}^{(2)}(2\mathbf{x}_i- \mathbf{x}_j- \mathbf{x}_p)]\, .
\end{align}%
\end{subequations}%
The corresponding synchronized solution is given by,
\begin{subequations}%
\begin{align}
\dot{\mathbf{s}} = & {f}(\mathbf{s}) + \sigma_1 r^{(1)}h^{(1)}(0) {g}^{(1)}(\mathbf{s},\mathbf{s}) \notag \\ & + 2\sigma_2r^{(2)}h^{(2)}(0) {g}^{(2)}(\mathbf{s},\mathbf{s},\mathbf{s}), \\ 
k_{ij}^{(s)} =& -a_{ij}^{(1)}{h}^{(1)}(0), \\ 
k_{ijp}^{(s)} =& -a_{ijp}^{(2)}{h}^{(2)}(0) \, .
\end{align}
A linear stability analysis about this solution can be performed to infer the existence of a synchronous solution. We thus perturb the system with small perturbation terms $\mathbf{\xi}_i = \mathbf{x}_i - \mathbf{s} $, $ \chi_{ij} = k_{ij}^{(1)} - k_{ij}^{(s)} $ and $\eta_{ijp} = k_{ijp}^{(2)} - k_{ijp}^{(s)}$, ans we study their time evolution via the variational equations %
\end{subequations}%
\begin{subequations}%
\begin{widetext}
\begin{align}
\label{5}
\dot{\mathbf{\xi}}_i = &  D{f}(\mathbf{s})\mathbf{\xi}_i +\sigma_{1}r^{(1)}h^{(1)}(0){D}{g}^{(1)}(\mathbf{s},\mathbf{s}) \xi_i  +2\sigma_{2}r^{(2)}h^{(2)}(0){D}{g}^{(2)}(\mathbf{s},\mathbf{s},\mathbf{s})\xi_{i} -\sigma_1 {h}^{(1)}(0)\sum_{j=1}^N {L}_{ij}^{(1)} {D}_2{g}^{(1)}(\mathbf{s},\mathbf{s})\xi_j +  \notag \\
& - \sigma_2{h}^{(2)}(0)\sum_{j=1}^N {L}_{ij}^{(2)}{D}_s {g}^{(2)}(\mathbf{s},\mathbf{s},\mathbf{s})\xi_j - \sigma_1 \sum_{j=1}^N a_{ij}^{(1)} \chi_{ij} {g}^{(1)}(\mathbf{s},\mathbf{s}) - \sigma_2\sum_{j=1}^N \sum_{p=1}^N a_{ijp}^{(2)}\eta_{ijp} {g}^{(2)}(\mathbf{s},\mathbf{s},\mathbf{s}) ,
\end{align}
\end{widetext}
\begin{align}
 \dot{\chi}_{ij} = & - \epsilon_1(\chi_{ij} + a_{ij}^{(1)}{Dh}^{(1)}(0)(\xi_i - \xi_j)), \\
 \dot{\eta}_{ijp} =& - \epsilon_2(\eta_{ijp} + a_{ijp}^{(2)}{Dh}^{(2)}(0)(2\xi_i - \xi_j - \xi_p))\, ,
\end{align}%
\end{subequations}%
where $D{g}^{(1)}=D_1{g}^{(1)}+ D_2{g}^{(1)}$, $D{g}^{(2)}=D_1{g}^{(2)}+ D_2{g}^{(2)} +  D_3{g}^{(2)}$ and $D_{s}{g}^{(2)}= D_2{g}^{(2)} +  D_3{g}^{(2)}$.

Now, we consider ${\xi} = \mathbf{x}- I_N \otimes \mathbf{s} $, $ {\chi} = k^{(1)} - k^s $ and $ {\eta} = k^{(2)} - k^s $ with
\begin{align}
& \mathbf{x} =  (\mathbf{x}_1^{\top}, \dots ,\mathbf{x}_N^{\top} )^{\top},  \notag \\   
& k^{(1)}  =  (k_{11}^{(1)},\dots,k_{1N}^{(1)},k_{21}^{(1)},\dots,k_{2N}^{(1)}, \dots, k_{N1}^{(1)},\dots, k_{NN}^{(1)})^{\top}, \notag \\
& k^{(2)}  =  (k_{111}^{(2)},\dots,k_{1NN}^{(1)},k_{211}^{(1)},\dots,k_{2NN}^{(1)}, \notag \\  & \hspace*{2cm}
\dots, k_{N11}^{(1)},\dots, k_{NNN}^{(1)})^{\top}.
\end{align}
Now analogous to the previous case here we again introduce few other notations and matrices to be used in the following derivations. We consider
\begin{align}
    \mathbf{a}_i^{(2)} = (a_{i11}^{(2)},a_{i12}^{(2)},\dots,a_{iNN}^{(2)}),  \notag 
\end{align}
and  $N \times N^3$, $N^3 \times N$ matrices as,
\begin{align}
\mathbf{B}^{(2)} & = 
\begin{pmatrix}
\mathbf{a}_{1}^{(2)}& & \\
    &\mathbf{a}_{2}^{(2)}& \\
     &\dots& \\
     & & \mathbf{a}_{N}^{(2)}  
\end{pmatrix},\notag   \\
\mathbf{C}^{(2)} & = 2\mathbf{B}^{(2)^{\top}} - \mathbf{D}^{(2)}_{1} - \mathbf{D}^{(2)}_{2},
\end{align}
where the  $N^3 \times N$ matrices $\mathbf{D}^{(2)}_{1}$ and $\mathbf{D}^{(2)}_{2}$ is given by 
\begin{align}
\mathbf{D}^{(2)}_{1} = 
\begin{pmatrix}
    {d}_1 \\
     \vdots \\
     {d}_N   
\end{pmatrix}, \;\;\;
\mathbf{D}^{(2)}_{2} = 
\begin{pmatrix}
    {e}_1 \\
     \vdots \\
     {e}_N  
\end{pmatrix}. \notag  
\end{align}
The elements of the matrices $\mathbf{D}_{1}^{(2)}$ and $\mathbf{D}_{2}^{(2)}$ are given for $i=1,2,\dots,N$ as
\begin{align}
	{d}_{i} = 
	\begin{pmatrix}
		p_{i1} & & \\
		& p_{i2} &\\
		& & \dots &\\
		& & &p_{iN}    
	\end{pmatrix},
\;\;
\text{and}
\;\;
	{e}_{i} = 
	\begin{pmatrix}
		q_{i1}\\
		q_{i2}\\
		\vdots \\
		q_{iN}  
	\end{pmatrix}\notag  
\end{align}
with
\begin{align}
{p}_{ij} = 
\begin{pmatrix}
    a_{ij1}^{(2)}\\
    a_{ij2}^{(2)}\\
     \vdots \\
    a_{ijN}^{(2)}     
\end{pmatrix},
\;\;
\text{and}
\;\;
{q}_{ij} = 
\begin{pmatrix}
    a_{ij1}^{(2)}& & \\
    &a_{ij2}^{(2)}&\\
    & & \dots &\\
    & & &a_{ijN}^{(2)}  
\end{pmatrix}, \notag
\end{align}
where $j=1,2,\dots,N$.
Using all these notations together, the variational equation in the block matrix form can be expressed as 
\begin{widetext}
\begin{align}
\begin{bmatrix}
        \dot{\pmb{\xi}} \\
        \dot{\pmb{\chi}} \\
        \dot{\pmb{\eta}}
\end{bmatrix} 
= 
\begin{bmatrix}
        \mathbf{S}_2 & -\sigma_1 \mathbf{B}^{(1)} \otimes {g}^{(1)}(\mathbf{s},\mathbf{s}) & -\sigma_2 \mathbf{B}^{(2)} \otimes {g}^{(2)}(\mathbf{s},\mathbf{s},\mathbf{s}) \\
        -\epsilon_{1} \mathbf{C}^{(1)} \otimes Dh^{(1)}(0) & -\epsilon_{1} \mathbf{I}_{N^2} & 0 \\
        -\epsilon_{2} \mathbf{C}^{(2)} \otimes Dh^{(2)}(0) & 0 & -\epsilon_{2} \mathbf{I}_{N^3} 
\end{bmatrix} 
\begin{bmatrix}
\pmb{ {\xi}} \\
\pmb{\chi} \\
\pmb{\eta}
\end{bmatrix} \, ,
\end{align}
\end{widetext}
where we consider again the matrix $\pmb{\chi}$, resp. the tensor $\pmb{\eta}$, as $N^2$, resp. $N^3$, columns vectors, and we introduce suitable matrix
\begin{align}
\mathbf{S}_2 = & \mathbf{I}_N \otimes D{f}(f{s}) + \sigma_1h^{(1)}(0)(r^{(1)}\mathbf{I}_N \otimes D {g}^{(1)}) \notag \\  
+ & 2\sigma_2h^{(1)}(0)( r^{(2)}\mathbf{I}_N \otimes D{g}^{(2)}) - \sigma_1h^{(1)}(0)\mathbf{L}^{(1)} \otimes D_{s}{g}^{(1)} \notag \\ 
- & \sigma_2h^{(2)}(0)\mathbf{L}^{(2)} \otimes D_{s}{g}^{(2)}.
\end{align}
$\mathbf{B}^{(1)}$ and $\mathbf{C}^{(1)}$ are the same constant matrices of order $N \times N^2$ and $N^2 \times N$ used in Section~\ref{sec:nohoiadaptsupply}, while $\mathbf{B}^{(2)}$ and $\mathbf{C}^{(2)}$ are constant matrices of order $N \times N^3$ and $N^3 \times N$, satisfying $\mathbf{B}^{(2)}\mathbf{B}^{(2)^{\top}}=2r^{(2)}\mathbf{I}_N$ and $\mathbf{B}^{(2)}\mathbf{C}^{(2)}=\mathbf{L}^{(2)}$. 

Now to move forward we consider $\epsilon_{1}=\epsilon_{2}=\epsilon$. Then, from the structure of the variational equation, it is obvious that it has $(N^3+N^2-2N)$ numbers of eigenvalues $\lambda=-\epsilon$. Thereafter proceeding similarly as the pairwise adaptation case, we can define two matrices $\mathbf{Q}^{(2)}$ and $\mathbf{R}^{(2)}$ as,  
\begin{align}
\mathbf{Q}^{(2)}
=
\begin{pmatrix}
           \mathbf{I}_{Nd} & 0 & 0 & 0\\
           0& (1/r^{(1)})\mathbf{B}^{(1)^{\top}} & 0 & 0 \\
           0 & 0 & (1/r^{(2)})\mathbf{B}^{(2)^{\top}} & \mathbf{R}^{(2)}
\end{pmatrix}
\end{align}
and the left inverse of $\mathbf{Q}^{(2)}$ is
\begin{align}
\mathbf{Q}^{(2)^{-1}}
=
\begin{pmatrix}
           \mathbf{I}_{Nd} & 0 & 0\\
           0& \mathbf{B}^{(1)} & 0\\
           0 & 0 & \mathbf{B}^{(2)} \\ 
           0& & 0 & \mathbf{R}^{{(2)}^{\top}}
\end{pmatrix}.
\end{align}
So that, $\mathbf{Q}^{(2)^{-1}}\mathbf{Q}^{(2)} = \mathbf{I}_{N^3+N^2+Nd}$. 
Then, the variational equation can be transformed as,
\begin{widetext}
\begin{align}
\begin{pmatrix}
        {\dot{\pmb{\xi}}} \\
        \dot{\pmb{\chi}} \\
        \dot{\pmb{\eta}}
\end{pmatrix} 
= 
\mathbf{Q}^{(2)^{-1}}
\begin{bmatrix}
        \mathbf{S}_2 & -\sigma_1 \mathbf{B}^{(1)} \otimes {g}^{(1)}(\mathbf{s},\mathbf{s}) & -\sigma_2 \mathbf{B}^{(2)} \otimes {g}^{(2)}(\mathbf{s},\mathbf{s},\mathbf{s}) \\
        -\epsilon \mathbf{C}^{(1)} \otimes Dh^{(1)}(0) & -\epsilon \mathbf{I}_{N^2} & 0 \\
        -\epsilon \mathbf{C}^{(2)} \otimes Dh^{(2)}(0) & 0 & -\epsilon \mathbf{I}_{N^3} 
\end{bmatrix} 
\mathbf{Q}^{(2)}
\begin{pmatrix}
        { \pmb{\xi}} \\
        \pmb{\chi} \\
        \pmb{\eta}
\end{pmatrix}, 
\end{align}
where again, with a slight abuse, we use the same letter to indicate the new transformed coordinates.
Now,
\begin{align}
\mathbf{Q}^{(2)^{-1}} &
\begin{bmatrix}
        \mathbf{S}_2 & -\sigma_1 \mathbf{B}^{(1)} \otimes {g}^{(1)}(\mathbf{s},\mathbf{s}) & -\sigma_2 \mathbf{B}^{(2)} \otimes {g}^{(2)}(\mathbf{s},\mathbf{s},\mathbf{s}) \\
        -\epsilon \mathbf{C}^{(1)} \otimes Dh^{(1)}(0) & -\epsilon \mathbf{I}_{N^2} & 0 \\
        -\epsilon {C}^{(2)} \otimes Dh^{(2)}(0) & 0 & -\epsilon \mathbf{I}_{N^3} 
\end{bmatrix} 
\mathbf{Q}^{(2)} \notag \\ 
& =
\mathbf{Q}^{(2)^{-1}}
\begin{bmatrix}
        \mathbf{S}_2 & -\sigma_1 \mathbf{I}_{N} \otimes {g}^{(1)}(\mathbf{s},\mathbf{s}) & -\sigma_2 \mathbf{I}_{N} \otimes {g}^{(2)}(\mathbf{s},\mathbf{s},\mathbf{s}) & 0\\
        -\epsilon \mathbf{C}^{(1)} \otimes Dh^{(1)}(0) & -\epsilon/r^{(1)} \mathbf{B}^{(1)^{\top}} & 0 & 0\\
        -\epsilon \mathbf{C}^{(2)} \otimes Dh^{(2)}(0) & 0 & -\epsilon/2r^{(2)} \mathbf{B}^{(2)^{\top}} & \epsilon \mathbf{R}^{(2)}
\end{bmatrix} \notag \\
& =
\begin{bmatrix}
        \mathbf{S}_2 & -\sigma_1 \mathbf{I}_{N} \otimes {g}^{(1)}(\mathbf{s},\mathbf{s}) & -\sigma_2 \mathbf{I}_{N} \otimes {g}^{(2)}(\mathbf{s},\mathbf{s},\mathbf{s}) & 0\\
        -\epsilon \mathbf{L}^{(1)} \otimes Dh^{(1)}(0) & -\epsilon \mathbf{I}_{N} & 0 & 0\\
        -\epsilon \mathbf{L}^{(2)} \otimes Dh^{(2)}(0) & 0 & -\epsilon \mathbf{I}_{N} & 0 \\
        -\epsilon \mathbf{R}^{(2)^{\top}}\mathbf{C}^{(2)} \otimes Dh^{(2)}(0) & 0 & 0 & -\epsilon \mathbf{I}_{N^3+N^2-2N}
\end{bmatrix}
\end{align}
From this, we can write $(Nd+2N)$ dimensional coupled master equations and $(N^3+N^2 - 2N)$ coupled salve equations as
\begin{align}
\begin{pmatrix}
        \dot{\pmb{\xi}} \\
        \dot{\pmb{\chi}}_M \\
        \dot{\pmb{\eta}}_M
\end{pmatrix} 
=
\begin{pmatrix}
\mathbf{S}_2 & -\sigma_1 \mathbf{I}_{N} \otimes {g}^{(1)}(\mathbf{s},\mathbf{s}) & -\sigma_2 \mathbf{I}_{N} \otimes {g}^{(2)}(\mathbf{s},\mathbf{s},\mathbf{s}) \\
        -\epsilon \mathbf{L}^{(1)} \otimes Dh^{(1)}(0) & -\epsilon \mathbf{I}_{N} & 0 \\
        -\epsilon \mathbf{L}^{(2)} \otimes Dh^{(2)}(0) & 0 & -\epsilon \mathbf{I}_{N}  \\
        \end{pmatrix} 
\begin{pmatrix}
        {\pmb{\xi}} \\
         \pmb{\chi}_M \\
         \pmb{\eta}_M        
\end{pmatrix}.
\end{align}
Here, $(\pmb{\chi}_M,\pmb{\eta}_M) = (\pmb{\chi}_1,\pmb{\eta}_1)$ and the slave equations read as,
\begin{align}
\begin{pmatrix}
    \dot{\pmb{\chi}}_S \\
        \dot{\pmb{\eta}}_S 
    \end{pmatrix}
    =
\begin{pmatrix}
     -\epsilon \mathbf{R}^{(2)^{\top}}\mathbf{C}^{(2)} \otimes Dh^{(2)}(0) & 0 & -\epsilon \mathbf{I}_{N^2-N} 0 & -\epsilon \mathbf{I}_{N^3-N}
\end{pmatrix}
\begin{pmatrix}
        {\pmb{\xi}} \\
        {\pmb{\chi}}_M \\
         {\pmb{\chi}_S} \\
         \pmb{\eta}_M \\
         \pmb{\eta}_S
\end{pmatrix},
\end{align}
\end{widetext}
where ${\pmb{\chi}}_S = ({{\chi}}_2^{\top}, \ldots ,{{\chi}}_N^{\top})^{\top}$ and ${{\eta}}_S = ({{\eta}}_2^{\top}, \ldots ,{{\eta}}_N^{\top})^{\top}$.

Now again, to decouple the coupled master variational equation, we introduce a similar unitary transform like the pairwise adaptation case and introduce the new sets of co-ordinates as,
\begin{align}
	\begin{pmatrix}
		\mathbf{U} \otimes \mathbf{I}_{d} & 0 & 0\\
		0 & \mathbf{U} & 0 \\
		0 & 0 & \mathbf{U}
	\end{pmatrix} \begin{pmatrix}
		\pmb{\xi} \\
		\pmb{\chi}_{M} \\
		\pmb{\eta}_{M}
	\end{pmatrix}= \begin{pmatrix}
		\hat{\pmb{\xi}} \\
		\hat{\pmb{\chi}} \\
		\hat{\pmb{\eta}}
	\end{pmatrix}, 
	\notag\end{align}
with $\mathbf{U}$ the matrix whose columns are the orthonormal eigenvectors that diagonalize the pairwise Laplacian $\mathbf{L}^{(1)}$, i.e., $\mathbf{U}^{\top}\mathbf{L}^{(1)}\mathbf{U}=\mathbf{D}=\mathrm{diag} \{\mu^{(1)}_{1},\mu^{(1)}_{2}, \dots, \mu^{(1)}_{N} \}$. 
In terms of the changed coordinate frame the variational equation becomes
\begin{widetext}
\begin{align}
	\begin{pmatrix}
		\dot{\hat{\pmb{\xi}}} \\
		\dot{\hat{\pmb{\chi}}} \\
		\dot{\hat{\pmb{\eta}}}
	\end{pmatrix} 
	=
	\begin{pmatrix}
		\tilde{\mathbf{S}}_2 & -\sigma_1 \mathbf{I}_{N} \otimes {g}^{(1)}(\mathbf{s},\mathbf{s}) & -\sigma_2 \mathbf{I}_{N} \otimes {g}^{(2)}(\mathbf{s},\mathbf{s},\mathbf{s}) \\
		-\epsilon \mathbf{D} \otimes Dh^{(1)}(0) & -\epsilon \mathbf{I}_{N} & 0 \\
		-\epsilon \tilde{\mathbf{L}}^{(2)} \otimes Dh^{(2)}(0) & 0 & -\epsilon \mathbf{I}_{N}  \\
	\end{pmatrix} 
	\begin{pmatrix}
	\hat{\pmb{\xi}} \\
	\hat{\pmb{\chi}} \\
	\hat{\pmb{\eta}}    
	\end{pmatrix},
\end{align}
where $\tilde{{\mathbf{L}}}^{(2)}=\mathbf{U}^{\top}\mathbf{L}^{(2)}\mathbf{U}$ and $\tilde{\mathbf{S}}_{2}=\mathbf{U}^{\top}\mathbf{S}_{2}\mathbf{U}$. In explicit form, it can be rewritten as %
\begin{subequations}%
\begin{align}
	\dot{\hat{\xi}}_{i} =& \left[D{f}(\mathbf{s}) + \sigma_1h^{(1)}(0)r^{(1)} D{g}^{(1)} + 2\sigma_2 r^{(2)}h^{(2)}(0) D{g}^{(2)} + \sigma_1h^{(1)}(0) \mu^{(1)}_{i} D_{s} {g}^{(1)} \right]\hat{\xi}_{i} -\sigma_{2}h^{(2)}(0)\sum\limits_{j=1}^{N} \Tilde{L}^{2}_{ij}D_{s} {g}^{(2)}\hat{\xi}_{j} \notag \\ 
	&  -\sigma_1{g}^{(1)}(s,s)\hat{\chi}_{i} -\sigma_2{g}^{(2)}(s,s,s)\hat{\eta}_{i} ,  \\
	\dot{\hat{\chi}}_{i} =& -\epsilon[ \mu^{(1)}_{i} Dh^{(1)}(0) \hat{\xi}_{i} + \hat{\chi}_{i} ], \\
	\dot{\hat{\eta}}_{i} =& -\epsilon[ \sum\limits_{j=1}^{N} \Tilde{L}^{2}_{ij} Dh^{(2)}(0)\hat{\xi}_{i}  + \hat{\eta}_{i} ]\, ,
	\label{genmse2B}
\end{align}
which is the master stability equation we need.   
\end{subequations}
\end{widetext}
\end{document}